\documentclass{appolb}
\usepackage{latexsym}
\usepackage{graphicx} 
\usepackage{epsfig}
\usepackage{amssymb} 
\newcommand{\MeV}{{\rm MeV}} 
 
\newcommand{\fm}{{\rm fm}}

\newcommand{\bquote}{\begin{quote}}
\newcommand{\equote}{\end{quote}}

\newcommand{\dsp}{\displaystyle}

\newcommand\Eqn[1]{Eq.~(\ref{#1})}  % includes ``Eq.'' in front
\newcommand{\beq}{\begin{equation}}
\newcommand{\eeq}{\end{equation}}
\newcommand{\ba}{\begin{array}}
\newcommand{\bea}{\begin{eqnarray}}
\newcommand{\ea}{\end{array}}
\newcommand{\eea}{\end{eqnarray}}

\newcommand\comment[1]{ \hbox{[{\it Comment suppressed here.}\/]} }
\newcommand\hide[1]{}
\newcommand{\skipover}[1]{}

\def\lsim{\mathrel{\rlap{
\lower3pt\hbox{\hskip-3pt$\sim$}}
\raise1pt\hbox{$<$}}}
\newcommand{\gsim}{\raisebox{-0.7ex}{$\stackrel{\textstyle >}{\sim}$ }}
\begin{document}
\eqsec \newcount\eLiNe\eLiNe=\inputlineno\advance\eLiNe by -1
\title{Novel Phases at High Density and their Roles in the Structure
and Evolution of Neutron Stars \thanks{Lectures presented at the
Zakopane Summer School on Flavor Dynamics, 2002} } \author{Sanjay Reddy
\address{Center for Theoretical Physics, M.I.T., 77 Massachusetts Ave,
MA 02139, USA \\ Theoretical Division, Los Alamos National Laboratory,
NM 87545. USA }}
\maketitle
\begin{abstract}
We present a pedagogic discussion on the role of novel phases of dense
baryonic matter in ``neutron'' stars. Qualitative aspects of the
physics that drives phase transitions and some of its astrophysical
consequences are discussed. Observable aspects of neutron star
structure and early evolution of the newly born neutron star are
discussed in some detail.
\end{abstract}

\section{Introduction}
Despite decades of study, the question - {\it What is the ground state
of dense baryonic matter?} - remains in the realm of theoretical
speculation. Advances in our understanding of strongly interacting
many-particle systems and QCD suggests numerous candidate states which
have competitive free energy.  Conventional many-body techniques,
including those of lattice gauge theory and non-relativistic
variational and Greens function methods, have yet to make definitive
statements regarding the phase structure of matter at densities of
relevance to neutron stars.  In these lectures we discuss a few novel
phases of matter and highlight some of their astrophysical
consequences.  They include (i) baryonic matter with hyperons, (ii)
kaon condensation and (iii) normal and superconducting quark matter.
The choice of topics presented is a synthesis of collaborative
research the author has undertaken in the past four years. In this
sense, the scope of the article is limited and no attempt is made to
present a comprehensive review. For a detailed review of assorted
topics relating to the interior of neutron stars see
Ref. \cite{Blaschke:uj}.

Terrestrial experiments, such as heavy-ion colliders, have been
employed to probe new phases of matter that are likely to occur at
high energy density. These experiments have proven to be useful in
exploring the high temperature region of the phase diagram. They have
had limited success in probing the high-density, low-temperature
phases of QCD. Neutron stars are best suited to probe this region and
thereby provide information that is complementary to terrestrial
efforts. The first half of the lectures provide a lowbrow overview of
plausible new phases inside neutron stars. In the latter part we
discuss how these new phases will affect observable properties of
neutron stars. In particular, we focus on the following topics to make
contact with observables: (i) mass and radius, (ii) the birth and
early evolution of newly born neutron stars in a supernova explosion,
as evidenced by the neutrinos they emit.

\section{Nucleonic Matter}
\label{nuclear}
The thermodynamic properties of a many-particle system consisting of
strongly interacting nucleons is difficult to calculate from first
principles, even if the underlying Hamiltonian were known. Over the
past several decades there have been numerous attempts to compute the
bulk properties of nuclear and neutron-rich matter. These include
microscopic many-body calculations using realistic nucleon-nucleon
potentials and phenomenological relativistic and non-relativistic
mean-field theories. The former approach starts with a potential that
provides a good description of the measured nucleon-nucleon scattering
data and uses variational or Greens function techniques to obtain the
thermodynamic properties of the many-particle system. In mean field
models the relation to nucleon-nucleon scattering is abandoned in
favor of a simpler form of the interaction whose parameters are
determined by fitting the model predictions to empirical properties of
bulk nuclear matter at nuclear saturation density.

Walecka's field theoretical model in which the nucleons interact with
$\omega$, $\rho$ and $\sigma$ mesons provides a reasonable description
of nuclei and nuclear matter \cite{Walecka:qa}. It has been
used extensively to study nuclear properties (for a review see
Ref. \cite{Serot:1984ey}). The model we discuss supplements the
original Walecka model with self-interactions between mesons and was
introduced by Boguta and Bodmer \cite{Boguta:xi}. The model Lagrangian
is given by
\begin{eqnarray}
{\cal L}_N \!=&& \overline{\Psi}_N \! \left( i\gamma^\mu
\partial_\mu-m_N^\ast
-g_{\omega N}\gamma^\mu V_\mu -g_{\rho N}\gamma^\mu
\vec{\tau}_N\cdot \vec{R}_\mu \!\right)\! \Psi_N \nonumber
\\
&&{} +\frac{1}{2}\partial_\mu \sigma
\partial^\mu\sigma-\frac{1}{2}m_\sigma^2\sigma^2-U(\sigma)-\frac{1}{4}
V_{\mu\nu}V^{\mu\nu} \nonumber 
\\ 
&&{} +\frac{1}{2}m_\omega^2V_\mu
V^\mu-\frac{1}{4}\vec{R}_{\mu\nu}
\cdot\vec{R}^{\mu\nu}+\frac{1}{2}m_\rho^2\vec{R}_\mu \cdot
\vec{R}^\mu,
\end{eqnarray}
where $\Psi_N$ is the nucleon field operator with $\vec{\tau}_N$ the
nucleon isospin operator. Further, $m^\ast_N = m_N-g_{\sigma N}\sigma$
is the nucleon effective mass. This mass is reduced from the free
nucleon mass $m_N$ by the scalar field $\sigma$, taken to have
$m_\sigma=600$~MeV. The scalar self-interaction term is given by
\begin{equation}
U(\sigma)= \frac{b}{3}m_N(g_{\sigma N}\sigma)^3 + \frac{c}{4}
(g_{\sigma N}\sigma)^4\ ,
\end{equation}
where $b$ and $c$ are dimensionless coupling constants. The kinetic
energy terms for the vector fields corresponding to the $\omega$ and $\rho$
mesons involve $V_{\mu\nu} = \partial_\mu V_\nu - \partial_\nu
V_\mu$, and $ \vec{R}_{\mu\nu} = \partial_\mu \vec{R}_\nu
-\partial_\nu \vec{R}_\mu $ respectively.

The five coupling constants, $g_{\sigma N}$, $g_{\omega N}$, $g_{\rho
N}$, $b$, and $c$, are chosen as in Ref.~\cite{GBOOK} to reproduce
five empirical properties of nuclear matter at saturation density: the
saturation density itself is $n_0 =0.16{\rm ~fm}^{-3}$; the binding
energy per nucleon is 16 MeV; the nuclear compression modulus is 240
MeV; the nucleon effective mass at saturation density is $0.78 m_N$;
and the symmetry energy\footnote{The energy that depends on the
difference in neutron and proton densities.} is 32.5 MeV.

The model is solved in the mean-field approximation, wherein only the
time component of the meson fields have nonzero expectation values.
The symbols $\sigma_0,\omega_0$ and $\rho_0$ below denote expectation
values that minimize the free energy. In the absence of neutrino
trapping, charge-neutral stellar matter can be characterized by the
baryon and electron chemical potentials since these are the only
conserved charges. Weak interaction equilibrium ensures that
$\mu_n-\mu_p=\mu_e$. The free energy density for such a system is
given by \cite{GBOOK}
\begin{eqnarray}
\Omega_{\rm nuclear}&(&\mu_B,\mu_e)= -P(\mu_B,\mu_e) \nonumber \\
&=&
\frac{1}{\pi^2}\left(
\int_0^{k_{Fn}}dk~k^2~(\epsilon_n(k) - \mu_n) +
\int_0^{k_{Fp}}dk~k^2~(\epsilon_p(k) - \mu_p) \right) \,\nonumber \\
&+&\frac{1}{2}\left(m_{\sigma}^2\sigma_0^2 -
m_{\omega}^2\omega_0^2-m_{\rho}^2\rho_0^2\right)+U(\sigma_0) -
\frac{\mu_e^4}{12\pi^2} \ ,
\label{omeganuc}
\end{eqnarray}
where $P(\mu_B,\mu_e)$ is the pressure, $\mu_B=\mu_n$ is the baryon
chemical potential and
\begin{eqnarray}
\epsilon_n(k) &=& 
  \sqrt{k^2+{m^\ast_N}^2} +g_{\omega N} \omega_0 - \half g_{\rho N}\rho_0
  \,,\\ \epsilon_p(k) &=& \sqrt{k^2+{m^\ast_N}^2} +g_{\omega N} \omega_0
  + \half g_{\rho N}\rho_0 \,,
\end{eqnarray}
are the neutron and proton single particle energies in the mean field
approximation.  The corresponding Fermi momenta $k_{Fn}$ and $k_{Fp}$,
which minimize the free energy at fixed baryon and electron chemical
potentials, are given by solving
\begin{eqnarray}
\label{walecka1}
\ba{rcl}
\epsilon_n(k_{Fn}) &=& \mu_n \ ,\\
\epsilon_p(k_{Fp}) &=& \mu_p \ .
\ea
\end{eqnarray}
Further, the mean field expectation values for the meson fields that 
minimize \Eqn{omeganuc} are given by
\begin{eqnarray}
\label{walecka2}
\ba{rcl}
m_\sigma^2  \sigma_0 &=& 
  \dsp  g_{\sigma N} \left(\rule[-0ex]{0em}{2ex}
  n_s(k_{Fn}) + n_s(k_{Fp})\right) 
  - \frac{dU}{d\sigma} \ , \\[1ex]
m_\omega^2  \omega_0 &=& g_{\omega N} \left(\rule[-0ex]{0em}{2ex}
  n_B(k_{Fn})+n_B(k_{Fp})\right) \ , \\[1ex]
m_\rho^2  \rho_0 &=&  \half g_\rho \left(\rule[-0ex]{0em}{2ex}
  n_B(k_{Fp})-n_B(k_{Fn})\right) \,,
\ea
\end{eqnarray}
where baryon number and scalar densities, $n_B$ and $n_s$ respectively, are 
given by
\begin{eqnarray} \ba{rcl} 
n_B(k_F) &=& 
\langle \Psi_N^\dagger \Psi_N \rangle =
\dsp \sum_{i=n,p}~\frac{1}{\pi^2} \int_0^{k_{Fi}} dk\, k^2 \ ,
\\
[2ex] 
n_s(k_F) &=& 
\langle \bar{\Psi}_N  \Psi_N \rangle =
\dsp 
\sum_{i=n,p}
\frac{1}{\pi^2} \int_0^{k_{Fi}} dk \, k^2
\frac{m^\ast_N}{\sqrt{k^2 + {m^\ast_N}^2}} \ .  
\ea 
\end{eqnarray}
Note that in \Eqn{omeganuc} the electron contribution to the free
energy has also been included. In bulk matter the condition $\partial
\Omega_{\rm nuclear}/\partial \mu_e=0$, which enforces electric-charge
neutrality, uniquely determines $\mu_e$. The magnitude and the density
dependence of the electron chemical potential is sensitive to the
value of the nuclear symmetry energy, parametrized in this model as
the strength of the isovector interaction.
\begin{figure}
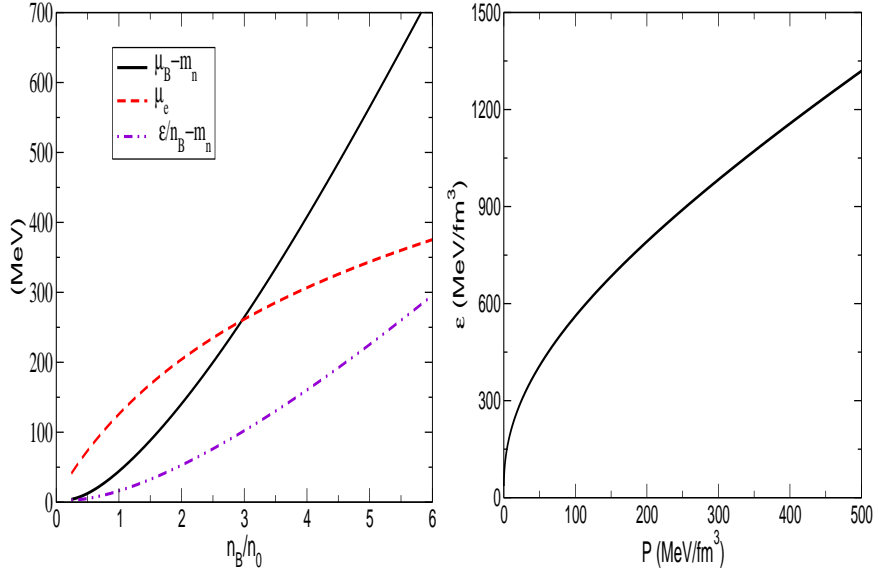

\includegraphics[width=0.45\textwidth,height=0.6\textwidth,angle=0]{nuceos1.eps}
\includegraphics[width=0.45\textwidth,height=0.6\textwidth,angle=0]{nuceos2.eps}
\caption{The nuclear equation of state. The left panel shows the
density dependence of the baryon chemical potential, the electron
chemical potential and the energy per baryon. The right panel shows
the relation between energy density and pressure.}
\label{nuceos}      
\end{figure}

The model described above provides us with a simple tool to understand
the various forces at play in the description of charge-neutral
nucleonic matter in weak equilibrium. To study possible phase
transitions we will employ this model to represent the nuclear phase.
Differences between the simple model employed here and detailed
many-body calculations can, however, be important at high
densities. This will influence both the location and the nature of the
phase transitions between the nuclear and more exotic phases to be
considered below. Nonetheless, the {\it qualitative} aspects of the
phase transitions are generic and are observed even in the most
sophisticated many-body descriptions of the nuclear phase.

Fig.~\ref{nuceos} shows the thermodynamic properties of charge-neutral
stellar matter calculated using the mean-field model.  The left panel
shows the baryon chemical potential, electron chemical potential and
the energy per particle as a function of the baryon density,
normalized to the nuclear saturation density $n_0=0.16$ fm$^{-3}$. The
right panel shows energy density as a function of pressure and is
usually referred to as the Equation of State (EoS).  It is this
relation that is required to solve for the structure of the neutron
star and therefore determines neutron star the mass and radius.

\section{Novel Phases}
With increasing density the chemical potentials for baryon number and
negative electric-charge increase rapidly due to the repulsive nature
of strong interactions at short distances. This furnishes energy for
the production of strange baryons and the condensation of mesons. At
even higher densities our knowledge of QCD and its asymptotic behavior
leads us to expect that the quarks inside nucleons will delocalize and form
a uniform Fermi sea of quarks. These expectations are borne out by
model calculations of these phases.  In what follows we consider a few
of these scenarios in detail.

\subsection{Hyperons}

Fig. 2 shows the variation of chemical potential associated with
neutral and charged baryons. The thick solid curves are predictions of
the mean field model described in the previous section and the thin
lines correspond to the non-interacting Fermi gas model.  The upper
curve in each of the above-mentioned cases corresponds to baryons with
negative charge, the middle curve to neutral baryons and the lowest
curve corresponds to baryons with positive charge. The upper and lower
horizontal dashed lines indicate the vacuum masses of the $\Sigma^\pm$
and the $\Lambda$ hyperons respectively. The location where the
chemical potentials cross the vacuum masses are also indicated for the
interacting case. This analysis neglects the strong interactions the
between hyperons and nucleons. If they were on-average attractive, the
corresponding second-order phase transitions would occur at lower
density. On the other hand, if they were repulsive, the transitions
would be pushed to higher densities. Note also that strong
interactions between nucleons also plays an important role in
determining the transition density. For the case of non-interacting
nucleons the transitions occur at densities that are significantly
larger. The baryon chemical potential in all models which incorporate
strong interactions will be larger than the naive prediction of the
free Fermi gas model at high densities. This generic feature arises
because strong interactions at short distances are repulsive.
\begin{figure}
\begin{center}
\includegraphics[width=0.8\textwidth,angle=-90]{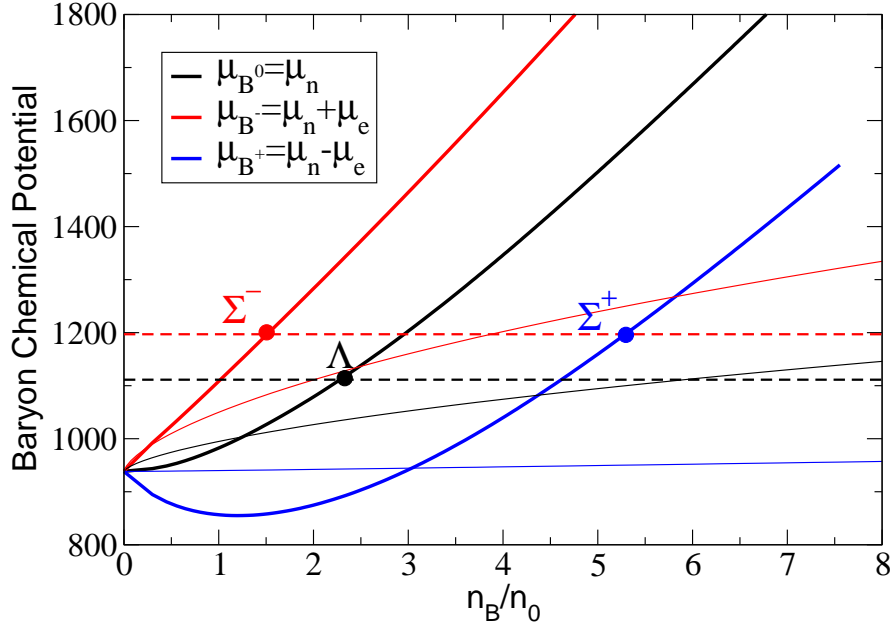}
\caption{Baryon chemical potentials in dense stellar matter computed
with (mean-field model, thick curves) and without (free Fermi gas
model, thin curves) strong interactions are shown. In each case, the
upper most curve is for negatively-charged baryons, the middle curve
is for neutral baryon and the lower curve is for positively-charged
baryons.}
\label{mub}   
\end{center}   
\end{figure}

These are second-order transitions because we are only considering the
energy cost for introducing infinitesimal hyperon populations. If the
hyperon-hyperon interactions are strongly attractive, these phase
transitions can become first-order. In this case it is energetically
favorable to have many hyperons appear together rather than for their
population to grow infinitesimally.

Hyperon-nucleon interactions are poorly constrained due to the lack of
experimental data. The exception is the $\Lambda$ binding energy in
hypernuclei \cite{Millener:hp}. This information was first employed in
mean field models by Glendenning \cite{Glendenning:1991es}. Hyperons
are introduced into the mean-field model discussed in section
\ref{nuclear} by minimal coupling to the $\sigma$, $\omega$ and $\rho$
mesons, and there arise three corresponding couplings constants that
must be constrained. The binding energy of the $\Lambda$ provides one
constraint which relates the $g_{H \sigma}$ and $g_{H \omega}$, since
the $\Lambda$ carries no isospin, and additionally the $\rho$ meson
mean field vanishes in isospin symmetric matter. In the simplest
mean-field model, the hyperon couplings to the vector meson couplings
are chosen to be similar to $2/3$ of that of the nucleons. This choice
is partially motivated by simple quark number and isospin counting
rules. The $\Lambda$ binding energy is then used to determine the
strength of the coupling to the $\sigma$ meson. The model further
assumes that the all hyperons couple to the mesons with the same
strength. This is likely to be a serious drawback of this simple
model. There are several direct and indirect indications coming
primarily from studies of $\Sigma^-$ atoms\footnote{This is an
electromagnetic bound state of $\Sigma^-$ and a nucleus} that indicate
the $\Sigma^-$ experiences a repulsive interaction in nuclear matter
\cite{Batty:zp}.

To summarize, strong interactions between nucleons result in a rapid
increase of the baryon chemical potential at high density. The baryon
chemical potential is typically larger than the lightest hyperon
masses at a baryon density between 1-3 times nuclear density.
Experimental inputs indicate that strong interactions between
$\Lambda$ particles and nucleons are attractive and this acts to lower
the density for their appearance. The situation with the $\Sigma^-$ is
less clear since experiment suggests that they experience repulsive
interactions in nuclei, even though the presence of a negative-charge
chemical potential in neutron stars favors their appearance.

Hyperons contribute more to the energy density of matter than to the
pressure as compared to nucleons. This is easily understood by noting
that they have larger masses and have smaller Fermi momenta. The
change in energy density per unit change in pressure is large relative
to the nuclear case, and this behavior is usually referred to as
softening of the EoS. As we will discuss in subsequent sections this
will act to reduce both the maximum mass and the radius of neutron
stars. Hyperons can also influence transport properties and the
thermal evolution of the star since they furnish new degrees of
freedom that are less degenerate than the nucleons, which can more
readily participate in dissipative processes.

\subsection{Kaon Condensation}
\begin{figure}
\includegraphics[width=0.8\textwidth,angle=-90]{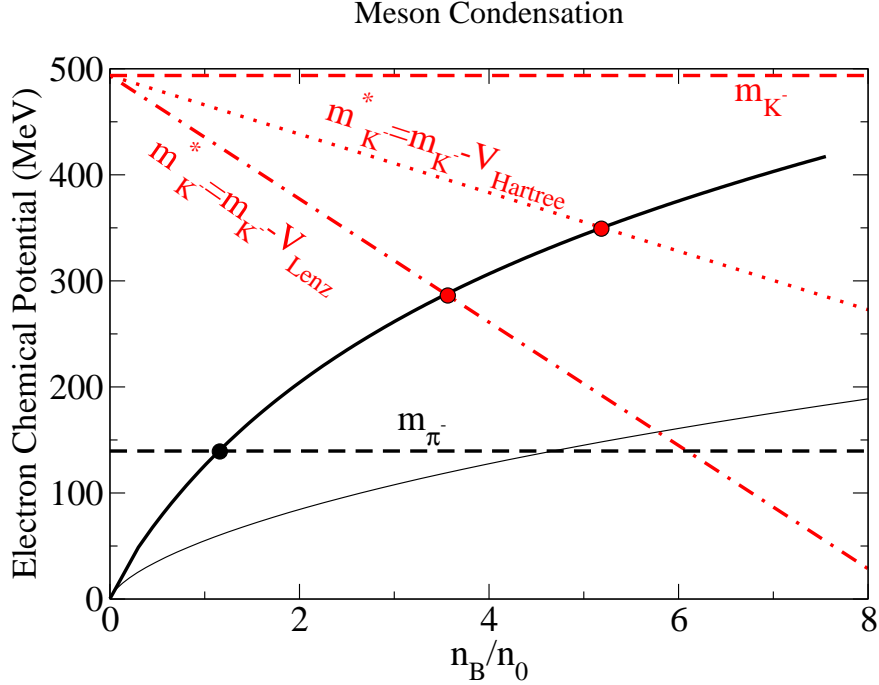}
\caption{Electron chemical potential, the pion and kaon vacuum and
in-medium effective masses in dense stellar matter}
\label{mue}      
\end{figure}

A large number of electrons are required to ensure charge neutrality
in dense nuclear matter. The typical electron chemical potential
$\mu_e$ is about $100$ MeV at nuclear density. With increasing
density, $\mu_e$ increases to keep pace with the increasing proton
number density. The magnitude of this increase depends sensitively on
the density dependence of the isovector interaction contribution to
the nuclear symmetry energy. In Fig.~\ref{mue} the density dependence
of the electron chemical potential in the mean-field model is shown as
the thick black curve.  For reference, the electron chemical potential
for the case of non-interacting nucleons is also shown (thin black
line). Negatively charged bosons whose rest energy in the medium is
less than the electron chemical potential will condense via a
second-order phase transition. The extent of condensation will be
regulated by repulsive interactions between bosons in the $s$-wave
channel at low momenta.

In the hadronic phase the likely candidates for condensation are the
$\pi^-$ and the $K^-$.  In vacuum, pions are significantly lighter
than kaons, but this situation may be reversed in the dense medium due
to strong interactions between mesons and nucleons. The physical basis
for this expectation is that the effective theory of meson-nucleon
interactions, such as chiral perturbation theory, must incorporate the
repulsive {\it s}-wave interactions arising due to the Pauli principle
between the constituent quarks inside nucleons and mesons. Mesons
containing $u$ or $d$ quarks will experience repulsion, while on the
other hand, mesons containing $\bar{u}$ or $\bar{d}$ quarks and
$s$ (i.e, kaons) quarks will experience attraction in the nuclear
medium.\footnote{This simple explanation is due to David Kaplan.}
Experiments with kaonic atoms lend strong support to the
aforementioned theoretical expectation of attraction
\cite{Friedman:hx}.

In Fig.~\ref{mue} the vacuum pion and kaon masses are shown as the
dashed lines. If the masses do not change in the medium, the figure
indicates that $\pi^-$ condensation occurs in the vicinity of nuclear
density and that $K^-$ condensation does not occur for the densities
considered. When interactions with medium are included, however, a
uniformly charged pion condensate is disfavored due to a weak
repulsive $s$-wave interaction. Instead, a spatially varying
condensate can be favored due to attractive $p$-wave interactions (for
a review see Ref. \cite{pionreview}). The kaon-nucleon interaction, on
the other hand, is {\it strongly} attractive. In what follows we only
discuss kaon condensation for this simple reason.

The idea that kaons could condense in dense nuclear matter was due to
Kaplan and Nelson \cite{Kaplan:yq}. Using a simplified $SU_R(3) \otimes
SU_L(3)$ chiral Lagrangian they showed that the $K^-$ could condense at a
density about three times nuclear density. Subsequently several
authors have studied in detail the nature and the role of kaon
condensation in neutron star matter (for a recent review see
Ref. \cite{Ramos:2000dq}).  

Here, we will employ a simple schematic potential model for
kaon-nucleon interactions considered in Ref. \cite{Carlson:1999rr} to
illustrate the salient features.  The scattering length $a_{K^- n}$
characterizes the low energy kaon-nucleon interaction, and experiment
indicates that $a_{K^- n} \approx -0.4$ fm. Following the authors of
Ref. \cite{Carlson:1999rr} we will assume that the interaction
potential can be modeled as a square well.  The parameters of the
potential, i.e, the depth $V_0=-122$ MeV and range $R=0.7$ fm, are
determined by fitting to low energy nucleon-kaon data (for further
details see Ref. \cite{Carlson:1999rr}). Given this potential the
change in the effective mass of $K^-$ in a low-density medium of
neutrons can be computed using the Lenz approximation. In this
approximation the attractive potential energy experienced by a kaon
at rest can be directly related to the scattering length and is given
by
\beq V_{\rm Lenz} = - \frac{2\pi}{m_{Kn}}~a_{K^- n}~n_n \,,
\label{lenz}
\eeq
where $m_{Kn}$ is the reduced mass of the neutron-kaon system and
$n_n$ is the neutron density. The effective mass of the kaon computed
using \Eqn{lenz} is shown in the Fig. \ref{mue}.  In this case, the
kaon effective mass equals the electron chemical potential when $n_n
\sim 3-4$ times nuclear density. This corresponds to the critical
density for kaon condensation.

At higher densities, the Hartree or mean field approximation is
expected to be valid. In this case the attractive potential energy of
the kaon cannot be related directly to on-shell low energy
kaon-nucleon scattering data. The Hartree potential is given by
\beq
V_{\rm Hartree} = \frac{4\pi}{3}~ V_0~R^3~n_n \,,
\label{hartree}
\eeq
where $V_0$ and $R$ are the depth and range of the $K^-$-$n$
potential.  Fig.~\ref{mue} indicates that kaons would condense in the
Hartree approximation when $n_n \sim 5-6$ times nuclear density. 

\subsection{Normal Quark Matter} 
\label{quark}
The occurrence of the novel hadronic phases mentioned above depends
sensitively on the nature of hadronic interactions and their many-body
descriptions. In contrast, the asymptotic behavior of QCD, which
requires that interactions between quarks become weak with increasing
momenta, provides strong support to the notion that at sufficiently
high densities nucleonic degrees of freedom must dissolve to form a
nearly free (perturbative) gas of quarks. The precise location of this
phase transition will depend on model descriptions of both the nuclear
and quark phases. This is because all model studies indicate that the
phase transition occurs at rather low densities where perturbative
methods do not apply.
\begin{figure}
\begin{center}
\includegraphics[width=0.8\textwidth]{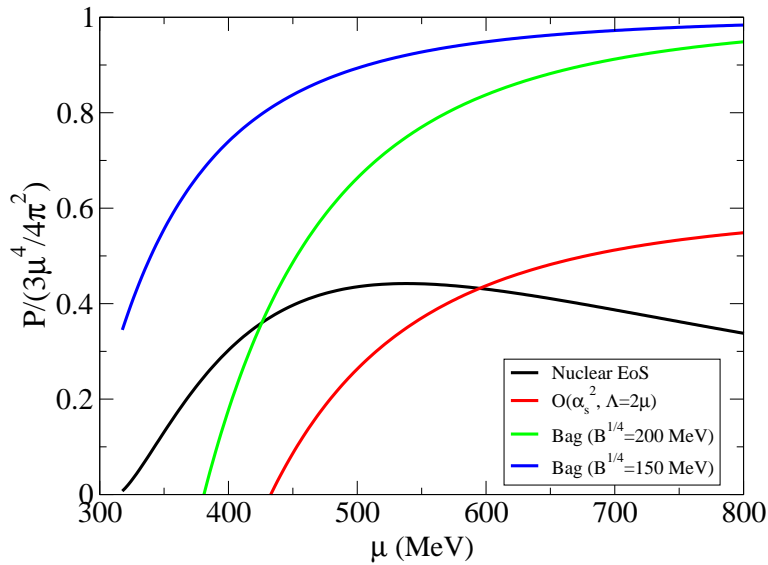}
\caption{Pressure v/s baryon chemical potential for several quark
matter EoS models. The pressure in the nuclear mean field model is
also shown for reference}
\label{qeos}  
\end{center}    
\end{figure}

The bag model provides a simple description of quark matter and
confinement. The model was designed to provide a description of the
hadron mass spectrum. The basic tenants of the model are a non-trivial
vacuum and nearly free quark propagation in spaces (bags) wherein the
perturbative vacuum has been restored. This restoration costs energy
since it requires the expulsion of the vacuum condensates. The
restoration energy per unit volume is called the bag constant and is
denoted as $B$. The model also provides a very simple and intuitive
description of bulk quark matter. The pressure in the bulk quark phase
containing up ($u$), down ($d$) and strange ($s$) quarks is due to the
kinetic energy density of quarks and a negative bag pressure. At zero
temperature this is given by
\beq P_{\rm Bag}(\mu_u,\mu_d,\mu_s) = - \Sigma_{i=u,d,s}
\int_0^{k_{Fi}}~\frac{\gamma~d^3k} {(2\pi)^3}(\sqrt{k^2+m_i^2} -\mu_i)
~ -B \,,
\label{pbag}
\eeq
where $\gamma=2_{\rm spins} \times 3_{\rm color}$ is the degeneracy
factor and $k_{Fi}$ is the Fermi momentum of each quark flavor. The
chemical potential for each flavor $\mu_i=\sqrt{ k_{Fi}^2 +
m_i^2}$, where $m_i$ is the corresponding quark mass. In the
limit of massless quarks and a common chemical potential for all the
quarks, the pressure in the bag model has the following simple form
\beq 
P_{\rm Bag}(\mu) = \frac{3}{4\pi^2} \mu^4 - B \,,
\label{masslessbag}
\eeq
where $\mu=\mu_B/3$ is the quark chemical potential.  The pressure of
bulk quark matter computed using the \Eqn{pbag} for $B^{1/4}=150$
MeV and $B^{1/4}=200$ MeV are shown in Fig.~\ref{qeos}. The phase with
the largest pressure is favored and the figure shows that the
nuclear-quark transition density is very sensitive to the bag
constant. For the case when $B^{1/4}= 150$ MeV (the upper most curve),
three flavor quark matter is the true ground state of matter and
nuclear matter is a metastable state \cite{sqm}. For larger values of
$B$ the transition occurs at higher densities.

Leading order effects of perturbative interactions between quarks can
also be incorporated in the bag model at high density. This has the
effect of renormalizing the kinetic term. The pressure in this case is
given by
\beq 
P_{{\rm bag},\alpha_s}(\mu) = \frac{3}{4\pi^2}~
\left(1+\frac{2~\alpha_s}{\pi}\right)~\mu^4 - B \,,
\eeq
where $\alpha_s=g^2/4\pi$ and $g$ is the QCD coupling.  At densities
of relevance to neutron stars, perturbative expansion in $\alpha_s$ is
not valid. Nonetheless, it is still interesting to note that the order
$\alpha_s^2$ calculation of the free energy predicts a behavior that
is similar to that of bag
model \cite{Freedman:1977gz,Fraga:2001id}. Recently, Fraga, Pisarski
and Schaffner-Bielich recomputed the equation of state of massless
quark matter to $O(\alpha_s^2)$. They find that the perturbative
result is well reproduced by the following bag-model-inspired form
for the pressure over a wide range of densities relevant to neutron
stars \cite{Fraga:2001id},
\beq 
P_{\rm perturb}(O(\alpha_s^2),\mu) = \frac{3}{4\pi} a_{\rm eff}
\mu^4 - B_{\rm eff} 
\eeq
\label{oa2}
where $a_{\rm eff} = 0.628$ and $B_{\rm eff}^{1/4}= 199$ MeV for the
specific choice for the renormalization subtraction point,
$\Lambda=2\mu$ . The pressure obtained using \Eqn{oa2} is also
shown in Fig.~\ref{qeos}. In the order $\alpha_s^2$ calculation, the
transition occurs at even higher density compared to the case when
$B^{1/4}=200$ MeV. 

To summarize, much like in the case of hyperons and kaons, quark
matter softens the equation of state. The softening is both due to a
larger number of degrees of freedom in the quark phase and the bag
constant which makes a negative contribution to the pressure and a
positive contribution to the energy density.

\subsection{Superconducting Quark Matter}
\label{superquark}
Since the early work of Bardeen, Cooper and Schrieffer it has been
well known that degenerate Fermi systems are unstable in the presence
of arbitrarily weak attractive interactions at the Fermi surface
\cite{Bardeen:1957mv}. The instability is resolved by the formation of
a Bose condensate of Cooper pairs. As is well known, for the case of
charged fermions, like electrons, this leads to superconductivity. In
analogy, the presence of attractive interactions between quarks will
lead to pairing and color superconductivity. This was realized several
decades ago in early work by B. Barrios and S. Frautschi
\cite{Barrois:1977xd}.  Recent realization that the typical
superconducting gaps in quark matter are larger than those predicted
in these early works has generated renewed interest. Model estimates
of the gap at densities of relevance to neutron stars suggest that
$\Delta \sim 100$ MeV when $\mu=400$ MeV \cite{BigGap}. Excellent
accounts of these recent findings can be in Ref. \cite{CFLReview}. In
what follows, we provide a brief introduction to the subject and
emphasize aspects that impact neutron star phenomenology.

It is simple to verify that the one-gluon-exchange (OGE) potential
between quarks is attractive in the antisymmetric (color-triplet)
channel, such as $(r_1b_2- b_1r_2)/\sqrt{2}$. The potential is given
by $V_{qq}^{A}=-2\alpha_s/3r$. Similarly, the OGE quark-quark
potential in the symmetric (color-sextet) channel is repulsive, and
the potential is $V_{qq}^{S} = \alpha_s/3r $. The attraction in the
triplet channel can result in $s$-wave pairing between quarks in spin
zero and spin one channels. Explicit calculations show that the
pairing energy, $\Delta$, is especially large for the spin-zero
case. This type of pairing can only occur between unlike flavors of
quarks to ensure that the diquark pair (which is a boson) has a
wavefunction that is symmetric.

For three massless flavors the condensation pattern that minimizes the
free energy is called the color-flavor-locked scheme
\cite{Alford:1998mk}. Pairing occurs between states that are on
opposite sides of the Fermi surface. For massless particles the
spin-zero state involves either only left and only right-handed
quarks. The non-zero condensates in this phase are given by
\begin{eqnarray}
\hspace{-0.2in}
\langle \psi^{i \alpha}_{\rm a,L} \psi^{j \beta}_{\rm b,L}
\epsilon_{ij}\rangle =- \langle \psi^{i \alpha}_{\rm a,R} \psi^{j
\beta}_{\rm b,R}\epsilon_{ij} \rangle &=&
\Delta~\epsilon^{\alpha \beta A}~\epsilon_{abA} 
= \Delta~(\delta^{\alpha}_a \delta^{\beta}_b
-\delta^{\alpha}_b \delta^{\beta}_a)
\end{eqnarray}
where $\alpha,\beta$ are color indices, $a,b$ are flavor indices, and
$i,j$ are spinor. The Levi-Civita symbols in spinor space, in flavor
space and in color space ensures that the pair has spin-zero, and is
anti-symmetric in flavor and color space respectively. The last equality
of this equation explicitly shows that the condensation locks color
and flavor indices.  

\begin{figure}
\begin{center}
\includegraphics[width=0.5\textwidth]{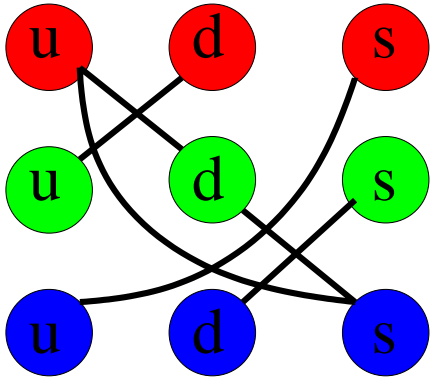}
\caption{Pairing scheme in color-flavor-locked quark matter. The
different rows (columns) correspond to different colors (flavors).}
\label{cflpair}
\end{center}      
\end{figure}

The choice of the colors and flavors that are locked, is of course,
arbitrary. For one particular choice the pairing scheme is illustrated
in Fig. \ref{cflpair}.\footnote{This picture illustration was suggested
by George Bertsch.}
In the color-flavor-locked phase (CFL henceforth) all nine quarks
participate in pairing. Consequently, the energy required to create a
quark excitation involves the breaking of a Cooper pair and the
fermion excitation spectrum is characterized by an energy gap $\sim 2
\Delta$. Further, gluons acquire a mass via the Higgs mechanism by
coupling to the colored condensates. The lowest energy collective
excitations in this phase correspond to Goldstone bosons. The number
and spectrum of these Goldstone modes can be understood by noting that
the condensate breaks baryon number and chiral symmetries.  Baryon
number is broken because the CFL state is not an eigenstate of the
baryon number operator. This is completely analogous to
superconducting state of electrons, where the BCS ground state is
not an eigenstate of the electron number.

The CFL state breaks chiral symmetry by a novel mechanism
\cite{Alford:1998mk}.  Left and right-handed condensates are coupled
because the color exchange is a vector interaction.  Flavor rotations
on the $L$ condensate must be compensated by a rotation of the $R$
condensate \cite{Alford:1998mk}, and therefore chiral symmetry is
spontaneously broken. The octet of flavor Goldstone bosons that result
will acquire a mass because quark masses explicitly break chiral
symmetry. The quantum numbers of these pseudo-Goldstone modes maps
onto the meson octet in vacuum. For this reason they are often
commonly referred to as the ``pions'' and ``kaons'' (or collectively
as ``mesons'') of the high-density phase. Their masses have been
computed in earlier work by matching the high density chiral effective
theory to perturbative QCD. Unlike the vacuum case, the square of
``meson'' masses here is proportional to the product of quark masses
\cite{CFLmesons}. This results in an inverted hierarchy, where the
pions are heavier than the kaons. Explicitly, at high density the
masses are given by
\begin{eqnarray}
m^2_{\pi^\pm} &=& a (m_u + m_d)m_s \cr
m^2_{K^\pm} &=& a (m_u + m_s)m_d \cr 
m^2_{K^0} &=& a (m_d + m_s)m_u \,, 
\label{masses} 
\end{eqnarray}
where $a=3\Delta^2 / \pi^2\mu^2$.  In contrast, the baryon number
Goldstone boson remains massless.  This singlet mode is responsible
for the superfluid nature of this phase \cite{Alford:1998mk}. As will
be discussed in detail in subsequent sections, this massless mode also
plays a crucial role in determining the neutrino opacity of the CFL
phase.

Pairing at the Fermi surface also contributes to the pressure, which on
dimensional grounds, we can expect to be $\sim \Delta^2 \mu^2$.  For
three massless flavors, we can supplement the bag model pressure with
the contribution due to superconductivity. In the CFL phase it is
given \cite{Alford:2001zr}
\beq
P_{\rm CFL} = \frac{3}{4\pi^2}~\mu^4 + \frac{3}{\pi^2}~\Delta^2 \mu^2 - B \,,
\eeq
which is to be compared with \Eqn{masslessbag}.

It is a reasonable approximation to neglect the $u$ and $d$ quark
masses at densities of relevance to neutron stars where $\mu \sim 400$
MeV.  The strange quark mass $m_s \sim 200$ MeV, on the other-hand,
cannot be neglected. The difference in Fermi momenta between light
quarks and the strange quark is $\sim m_s^2/2\mu$. Thus, when $\Delta
\lsim m_s^2/2\mu$, pairing involving strange quarks will be
suppressed. In the limit of infinite strange quark mass, i.e, in their
absence, only light quarks pair. This phase is called 2SC (two-flavor
superconductor) and it is also characterized by pairs that are
antisymmetric in flavor \cite{BigGap}.  Antisymmetry in color space
excludes one of the three colors from participating in the
condensation. Thus $SU(3)_{\rm color}$ is broken down to $SU(2)_{\rm
color}$. Quarks of a particular color and three gluons remain
massless. Further, unlike the CFL phase, no global symmetries are
broken. The absence of the massless Goldstone bosons implies the
absence of the superfluidity. In this sense, the 2SC phase appears
quite unremarkable.

Early attempts to bridge the ``gap'' between the 2SC phase and the CFL
phase can be found in Ref. \cite{Alford:1999pa,Buballa:2001gj}. They
found that the CFL pairing scheme is preserved when $m_s \lsim \sqrt{2
\mu \Delta}$. For larger $m_s$ a first-order transition to the 2SC
phase occurs.  The response of the CFL phase to a finite strange quark
mass in these earlier works ignored the role of the flavor Goldstone
modes. Bedaque and Schafer showed that the strange quark mass appeared
in the effective theory for Goldstone bosons in the form of a chemical
potential for anti-strangeness \cite{Bedaque:2001je}. Since $m_u<m_d$,
$K^0$ is lightest Goldstone mode with anti-strangeness in the
high-density effective theory (see \Eqn{masses}). Consequently, when
the chemical potential, $\mu_{\bar{s}} \sim m_s^2/2\mu$, exceeds the
mass of the $K^0$ boson they will condense in the ground state.  Using
the asymptotic formula for the meson masses given in \Eqn{masses},
Bedaque and Schafer showed that $K^0$ condensation occurs when $m_s
\gsim (m_u \Delta^2)^{1/3}$. The phase with a $K^0$
condensate is distinct from the the CFL phase as it breaks additional
symmetries. However, since the pairing scheme itself remains unaltered
we label this phase CFL$K^0$ \cite{Kaplan:2001qk}. In the presence of
quark masses and chemical potentials, the CFL phase is symmetric under
$U(1)_Y \otimes U(1)_{\tilde{Q}}$ rotations, where $Y$ is hypercharge
or strangeness and $\tilde{Q}$ is the modified electric-charge in the
CFL phase. In the CFL$K^0$ phase hypercharge or strangeness symmetry
is spontaneously broken, resulting in the appearance of a massless
Goldstone boson.

Much like the strange-quark mass, the presence of an electric-charge
chemical potential will act as source to condense charged mesons in
the CFL phase if its strength exceeds the mass of the lightest charged
boson. In the CFL phase, $K^+$ and $\pi^-$ are the lowest energy
charged-particle excitations. (The $K^-$ excitations cost more energy
due to the presence of the anti-strange chemical potential induced by
the strange quark mass.)  The phase diagram in the presence of an electric 
charge chemical potential and the strange quark mass exhibits a rich
structure.  For a detailed discussion of these novel meson condensed
phases and their roles in the birth and evolution of neutron stars see
Ref. \cite{Kaplan:2001qk}.

An interesting feature of BCS-like pairing in simple systems with only
two degrees of freedom is the locking of the Fermi surfaces in
momentum space. Consider pairing between massless $u$ and $d$ quarks
in the presence of an electric-charge chemical potential. The charge
chemical potential in the absence of the pairing interaction will act
to change the relative number of $u$ and $d$ quarks. In the presence
of pairing the state resists this change. With increasing
electric-charge chemical potential the system exhibits no response up
to a critical value. When the difference in chemical potentials
between the particles participating in the pairing exceeds $2 \Delta$,
the system responds via a first-order transition to the unpaired normal
state. This rigidity was initially thought to ensure equal numbers of
$u,d$ and $s$ quarks in the CFL phase, despite the strange quark mass
\cite{Rajagopal:2000ff}. A detailed analysis, however, shows that
rigidity in the color-flavor-locked phase takes on a different form
\cite{Steiner:2002gx}. It enforces equality between the color and
flavor quark number of those colors and flavors that are locked by
pairing. For the pairing scheme shown in Fig. \ref{cflpair} rigidity
manifests in the form of the following equations \cite{Steiner:2002gx}
\beq
n_{\rm r}=n_{\rm u} \qquad
n_{\rm g}=n_{\rm d} \qquad
n_{\rm b}=n_{\rm s}  \,.
\eeq
The above relations, combined with the additional condition of local
color neutrality, ensure equal numbers of $u$, $d$ and $s$
quarks \cite{Steiner:2002gx}. Therefore, despite the strange quark
mass, the CFL and the CFL$K^0$ phases are electrically neutral. No
electrons are present and there are no massless charged-particle
excitations. Thus, with regard to electromagnetism, the CFL phase is a
transparent insulator \cite{Manuel:2001mx}.

\section{Nature of the Phase Transition}
The novel phases discussed above could occur either via a first or
second-order phase transition. In the latter case, the order parameter
for the new phase is infinitesimal at the critical density and
smoothly grows with increasing density. In these situations the
electric-charge chemical potential is also able to adjust smoothly to
maintain charge neutrality in the bulk phase. On the other hand, in
the case of first-order phase transitions the order parameter
characterizing the new phase has a finite value at the onset of the
transition. The electric-charge chemical potential required to ensure
neutrality changes discontinuously across the transition. In these
situations the presence of two conserved charges, namely baryon number
and electric charge, allows for the possibility of phase co-existence,
i.e., satisfies the Gibbs criterion of equal chemical potentials and
pressure.

This phase with two co-existing phases is often referred to as the
mixed phase. Charge neutrality is enforced globally. The two
co-existing phases, in our case the nuclear phase and the denser
exotic phase, have opposite electric-charges, and carry positive and
negative charge respectively. The volume fraction occupied by the
exotic phase adjusts to ensure overall electric-charge
neutrality. Denoting the volume fraction of the denser phase as
$\chi$, global charge neutrality requires
\begin{equation}
\chi~Q_{\rm exotic} + (1-\chi)~Q_{\rm nuclear}=0 \,.
\label{chi}
\end{equation}
This was first noted by Glendenning \cite{Glendenning:1992vb} in the
context of high density phase transitions. The phase co-existence
occurs over a finite interval in pressure due to the presence of the
extra degree of freedom, namely electric-charge. However, as we
discuss next, the extent and existence of such a mixed phase depends
on surface tension between the two phases.

\begin{figure}[t]
\begin{center}
\includegraphics[width=0.8\textwidth]{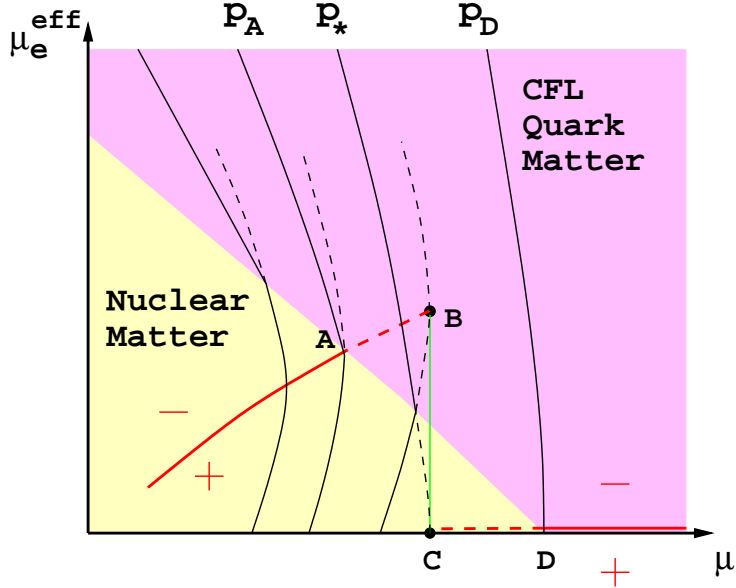}
\caption{A schematic phase diagram showing the nuclear and CFL phases
in the plane of quark chemical potential $\mu$ and effective electron
chemical potential $\mu_e$.  Isobars are shown as thin solid lines,
and neutrality lines for nuclear and CFL matter are thick lines.
Each phase is negatively charged above its neutrality line, and
positively charged below it.  Continuation onto the unfavored sheet is
shown by broken lines \cite{Alford:2001zr}.} 
\label{two_chem}
\end{center}
\end{figure}

To illustrate the physics of first-order phase transitions and the
role of surface and Coulomb energies in the mixed phase we consider an
explicit example. The phase transition from nuclear matter to CFL
quark matter is first-order. The nuclear phase has no strangeness and
the bulk quark phase has no electrons. The possibility of phase
co-existence between these phases was investigated in
Ref. \cite{Alford:2001zr}. We highlight some of the main findings
here. A schematic phase diagram is shown in Fig.~\ref{two_chem}. The
nuclear and CFL phases are favored in the lower-left and upper-right
shaded regions, respectively. With increasing baryon chemical
potential, the electron chemical potential in the charge-neutral
nuclear phase grows, as shown by the solid curve that ends at the
point labeled $A$. Nuclear matter is positively charge in the region
below this curve and negatively charged in the region
above. Similarly, the CFL phase is electrically neutral when
$\mu_e=0$, and is negatively (positively) charged when $\mu_e>0$
($\mu_e< 0$). When $\mu_e$ exceeds the mass of the lightest negatively
charged Goldstone mode these modes condense (via a second-order phase
transition) to provide an additional contribution to the
electric-charge density \cite{Alford:2001zr}. A positively charged
nuclear phase and a negatively charged CFL phase can coexist along the
line formed by the intersection of the two shaded regions (from
$A$-$D$). Along this line, which defines the mixed phase, both phases
have equal pressure and chemical potentials, i.e., they satisfy the
Gibbs criteria.

\begin{figure}[t]
\centering
{
\epsfig{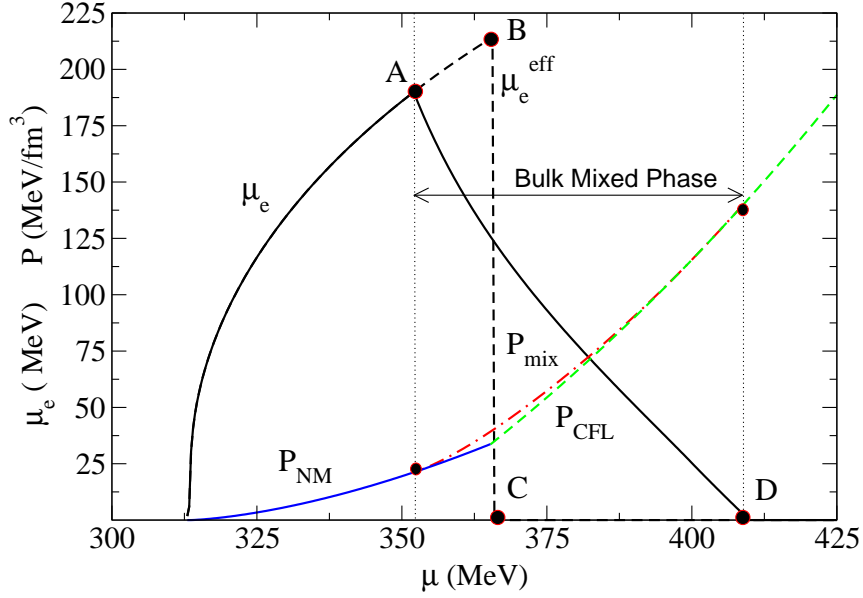}
}
\caption{Behavior of the electron chemical potential and the pressure
of homogeneous neutral nuclear matter, CFL matter and of the mixed
phase, all as a function of the quark chemical potential $\mu$. Only
bulk free energy is included; surface and Coulomb energy is neglected.
The mixed phase occurs between $A$ and $D$.  The vertical line
connecting $B$ and $C$ denotes the $\mu$ at which the pressures of
neutral CFL and nuclear matter are equal.  This is where a sharp
interface may occur.  The pressure of the mixed phase exceeds that of
neutral CFL or neutral nuclear matter between $A$ and
$D$ \cite{Alford:2001zr}.  }
\label{bigpicture}
\end{figure}

In Fig.~\ref{bigpicture}, a more detailed version of
Fig.~\ref{two_chem}, the pressure of the bulk nuclear, bulk CFL and
mixed phases are shown as a function of $\mu$, the quark chemical
potential. At intermediate values of $\mu$, the mixed phase has larger
pressure and is therefore favored over both the nuclear and CFL bulk
phases. The electron chemical potential, $\mu_e$, required to ensure
charge neutrality in the bulk nuclear phase, grows with $\mu$ as
shown. In the mixed phase neutrality requires a positively-charged
nuclear phase and a negatively-charged CFL phase. This is easily
accomplished by lowering $\mu_e$ from that required to maintain charge
neutrality in the nuclear phase (see Fig. \ref{two_chem}). The rate at
which $\mu_e$ decreases in the mixed phase is shown in the figure and
is obtained by requiring equal pressures in the both phases at a given
$\mu$.

As noted earlier, in the mixed phase the Coulomb and surface energy
costs must be met. The results shown in Fig.~\ref{bigpicture} ignore
these corrections. In a simple description of the mixed phase one
considers a sharp boundary between the coexisting phases. A unit cell
of the mixed phase is defined as the minimum-size region that is
electrically neutral. Three different simple geometries are
considered: spheres, rods and slabs \cite{pasta}. In each of these
cases, the surface and Coulomb energy cost per unit volume are given
by
\begin{eqnarray}
E^S &=& \frac{d~x~\sigma_{\rm QCD}}{r_0} \,, \label{surface}\\
E^C &=& 2\pi~\alpha_{\rm em}f_d(x)~(\Delta Q)^2~r_0^2 \,, 
\label{coulomb}
\end{eqnarray}
where $d$ is the dimensionality of the structure ($d=1,2,$ and $3$ for
slab, rod and droplet configurations, respectively), $\sigma$ is the
surface tension, and $\Delta Q=Q_{\rm nuclear}-Q_{\rm CFL+kaons}$ is
the charge-density contrast between the two phases. The other factors
appearing in Eqs.(\ref{surface}-\ref{coulomb}) are: $x$, the fraction
of the denser phase; $r_0$, the radius of the rarer phase (radius of
drops or rods and half-thickness of slabs); and $f_d(x)$, the
geometrical factor that arises in the calculation of the Coulomb
energy which can be written as\cite{pasta}
\begin{equation}
f_d(x)=\frac{1}{d+2}~\left(\frac{2-d~x^{1-2/d}}{d-2} + x\right) \ .
\end{equation}
The first step in the calculation is to evaluate 
$r_0$ by minimizing the sum of $E^C$ and $E^S$. The result is
\begin{equation}
r_0 = \left[\frac{d~x~\sigma_{\rm QCD}}{4\pi~\alpha_{\rm
em}f_d(x)~(\Delta Q)^2}\right]^{1/3} \,.
\label{radius}
\end{equation}
We then use this value of $r_0$ in Eqs. (\ref{surface}-\ref{coulomb}) to
evaluate the surface and Coulomb energy cost per unit volume
\begin{equation}
E^S+E^C = \frac{3}{2} \left(4\pi~\alpha_{em}~d^2~f_d(x)~x^2\right)^{1/3}
~(\Delta Q)^{2/3}~\sigma_{\rm QCD}^{2/3} \,,
\label{sandccost}
\end{equation}
where $x$ equals $\chi$ when $\chi \leqslant 0.5$ and $(1-\chi)$ when $0.5
\leqslant \chi \leqslant 1$. The dependence $\chi(\mu)$ is computed
using \Eqn{chi}.  

We must now compare this cost to the bulk free energy benefit of the
mixed phase.
\begin{figure}[t]
\centering { \epsfig{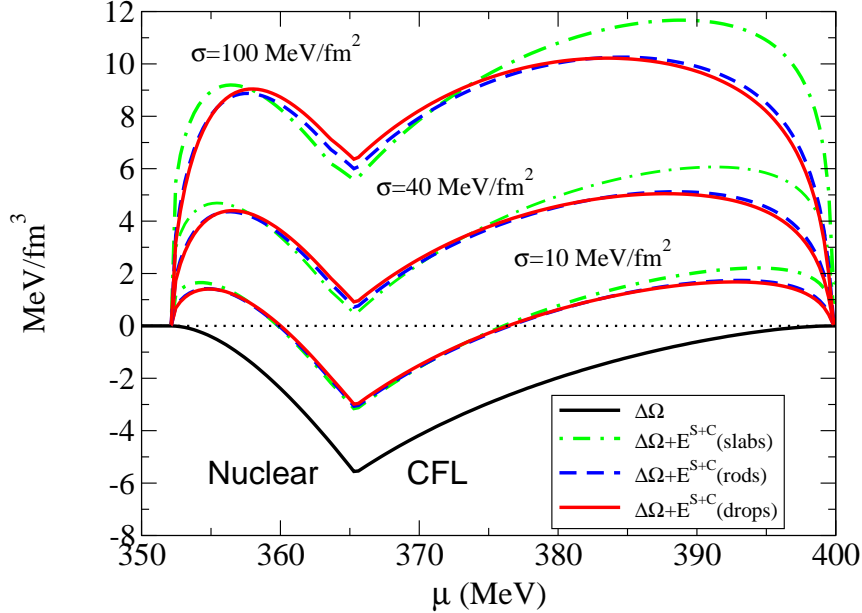} }
\caption{The free energy difference between the mixed phase and the 
homogeneous neutral nuclear and CFL phases. In the lowest curve,
the surface and Coulomb energy costs of the mixed phase
are neglected, and the mixed phase therefore 
has the lower free energy. Other curves include surface
and Coulomb energy for different values of $\sigma_{\rm QCD}$ and 
different mixed phase geometry.
As $\sigma_{\rm QCD}$ increases, 
the surface and Coulomb price paid by the mixed
phase increases.} 
\label{deltaomega}
\end{figure}
The lowest curve in Fig.~\ref{deltaomega} shows $\Delta \Omega$, the
difference between the free energy density of the mixed phase
(calculated without the surface and Coulomb energy cost) and the
homogeneous electrically neutral nuclear and CFL phases separated by a
single sharp interface, whichever of the two is lower. The mixed phase
has lower bulk free energy, so $\Delta\Omega$, plotted in
Fig.~\ref{deltaomega}, is negative.  The remaining curves in
Fig.~\ref{deltaomega} show the sum of the bulk free energy difference
$\Delta\Omega$ and $(E^S+E^C)$, the surface and Coulomb energy cost of
the mixed phase calculated using \Eqn{sandccost} for droplets, rods
and slabs. We employ three different values of $\sigma_{\rm QCD}$
since the QCD surface tension is, as yet, not very well known. Careful
inspection of the figure reveals that for any value of $\sigma_{\rm
QCD}$, the mixed phase is described as a function of increasing
density by a progression from drops to rods to slabs of CFL matter
within nuclear matter followed by slabs to rods to drops of nuclear
matter within CFL matter.  This is the same progression of geometries
seen in the inner crust of a neutron star \cite{pasta} or in the mixed
phase at a first-order phase transition between nuclear matter and
unpaired quark matter \cite{Heiselberg:1993dx} or hadronic kaon
condensate \cite{Glendenning:1998zx}.  We have also checked that for
$\sigma_{\rm QCD}=10$ and $40$~MeV/fm$^2$, with the mixed phase
geometry at any $\chi$ taken to be that favored, the sizes of regions
of both phases ($r_0$ and its suitably defined counterpart) are always
$\lsim 6$ fm. In general, uniform regions of charge can only exist on
length scales small compared to the Debye screening length. The Debye
screening length in the quark and hadronic phase are typically of the
size $5$ to $10$ fm \cite{Heiselberg:1993dx}.  When the size of the
charged regions becomes comparable to the Debye screening length, it
becomes important to account for spatial variations of the charge
density. This will influence the surface and Coulomb energy estimates
presented in \Eqn{sandccost}. A detailed discussion of the
importance of these finite-size effects is presented in
Ref. \cite{Norsen:2000wb}.

For any given $\sigma_{\rm QCD}$, the mixed phase has lower free
energy than homogeneous neutral CFL or nuclear matter wherever one of
the curves in Fig.~\ref{deltaomega} for that $\sigma_{\rm QCD}$ is
negative.  We see that some of the mixed phase will survive if
$\sigma_{\rm QCD}\approx 10~\MeV/ \fm^2$ while for $\sigma_{\rm QCD}
\gtrsim 40~\MeV/\fm^2$ the mixed phase is not favored for any $\mu$.
This means that if the QCD-scale surface tension $\sigma_{\rm QCD}
\gtrsim 40~\MeV/\fm^2$, a single sharp interface will be favored. The
interface is characterized by a bipolar charge distribution, resulting
in an intense electric field which ensures that the electric-charge
chemical potential is continuous across it (see
Ref. \cite{Alford:2001zr} for details).
\section{Mass and Radius of Neutron Stars}
The equations that enforce the condition of hydrostatic equilibrium in
compact stars where general relativistic effects are important are
called the Tolman-Oppenheimer-Volkov (TOV) equations
\cite{Oppenheimer:1939ne}.  They are given by
\begin{eqnarray}
\frac{dP}{dr}&=&\frac{-G~M(r)~\epsilon(P)}{r^2~c^2}
\left(1+\frac{P}{\epsilon}\right)
~\left(1+\frac{4\pi r^3 P}{M(r)c^2}\right)
\left((1-\frac{2G M(r)}{r c^2}\right)^{-1} \nonumber \\
\frac{dM(r)}{dr}&=&4\pi^2~\epsilon(P)
\end{eqnarray}
where $P=P(r)$ and the equation of state specifies $\epsilon(P)$,
i.e., the energy density as a function of the pressure, and $M(r)$ is
the total energy enclosed within radius $r$. For a given central
pressure, $P(r=0)$, the above equations can be easily integrated out
to the surface of the star, where $P=0$, to obtain the mass and radius
of the object.

\begin{figure}
\begin{center}
\includegraphics[width=0.8\textwidth,angle=-90]{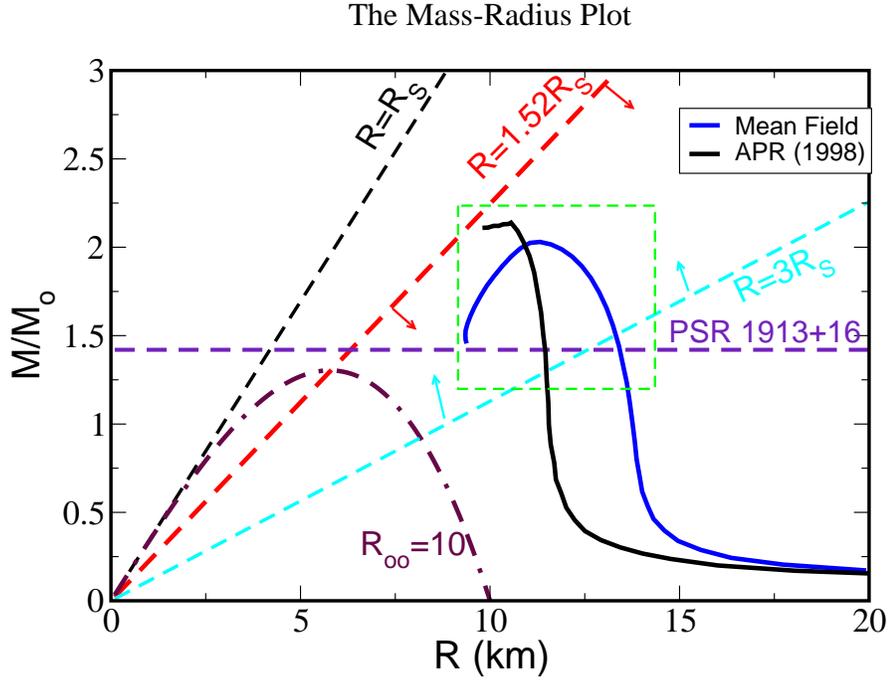}
\caption{Mass-radius relation for nuclear stars, where $M_o$ is the
mass of our sun.}
\label{masradiusnuc}      
\end{center}
\end{figure}

A family of stars all described by the same EoS can be obtained by
parametrically varying the central pressure and repeating the
procedure described above. Fig.~\ref{masradiusnuc} shows the
mass-radius curve obtained in this way for two different nuclear
equation of states. The curve labeled ``Mean-Field'' corresponds to
the model described in section \ref{nuclear} and the curve labeled APR
is a microscopic many-body calculation, by Akmal, Pandharipande and
Ravenhall, of the EoS using realistic nucleon-nucleon potentials
\cite{Akmal:1998cf}. The dashed lines labeled $R=R_s$ and $R=1.52R_s$
(where $R_s=2GM/c^2$ is Schwarzchild radius) are model-independent
constraints that require all stable stars to be to their right. To the
left of the line defined by $R=R_s$ the objects must be black holes
since the event horizon is outside the object, to the left of the
line defined by $R=1.52R_s$ the star requires a very stiff EoS for which
the speed of sound, $c_s=\sqrt{dP/d\epsilon}$,exceeds the speed of
light \cite{Weinberg}. The last stable orbit around a neutron star is
located at a radius $R=3R_s$. In some accreting neutron star systems
called QPO's (quasi-periodic oscillators) there is observational
evidence of the existence of such orbits. This implies that they exist
outside the physical radius of the star. This constrains stars to lie
to the left of the line defined by $R=3R_s$. Finally, the accurate measurement
of the neutron star mass in the Hulse-Taylor binary system (PSR
1913+16) introduces a further constraint which requires that the maximum
mass of the stars constructed using any model EoS be at least as
large as $1.44$ M$_\odot$ \cite{Hulse:1974eb}.

The radius of the neutron star as inferred from a point far away
(compared to $R_s$) is $R_\infty= R/\sqrt{1-R_s/R}$ where $R$ is the
radius of the star, as inferred from the TOV equation. An accurate
determination of $R_\infty$ would greatly constrain the allowed region
in the mass radius diagram. For example, a precise measurement of an
$R_\infty$ of 10 km constrains all stars to lie on the curve labeled
$R_\infty=10 $ km in Fig. \ref{masradiusnuc}. The discovery of a
nearby isolated neutron star which appears to emit nearly black-body
radiation promises to provide valuable information about $R_\infty$
\cite{Walter:nature}. Assuming that the object is emitting black-body
radiation, a measurement of the photon energy flux $F$ and the distance
to the object $d$ uniquely determine $R_\infty$ as
\begin{equation}
R_{\infty} = d~\sqrt{\frac{F}{\sigma_{\rm SB} T_{\infty}^4}}\,,
\end{equation}
where $\sigma_{\rm SB}$ is the Stefan-Boltzmann constant and
$T_{\infty}$ is the temperature of the black-body spectrum inferred at
infinity. In the case of RXJ 185635-3754, the x-ray spectrum is fit to
good approximation by a black-body spectrum. The distance to this
object $d\simeq 117$ pc \cite{Kaplan:2001vn,Walter:2002uq}. However,
the optical and x-ray data cannot be reconciled with a single
temperature black-body spectrum \cite{Pons:2001px}. Consequently a
stringent constraint of the form described by the $R_{\infty}=10$ km
curve is not yet possible \cite{Pons:2001px,Walter:2002uq}.

Phase transitions at high density generically result in an increase in
the energy density of matter at a given pressure. We will loosely refer
to this as softening. An EoS whose energy density (as a function of
pressure) is on-average large compared to the nuclear EoS can be
considered to be soft. A larger energy density increases the
gravitation attraction inside the compact object. Consequently, it
will require higher central pressures to achieve hydrostatic
equilibrium for a given mass of the compact object. However, with
increasing pressure, the energy density increases rapidly in a soft
EoS. The competition between matter pressure and gravity is heavily
stacked in favor of gravity in a star with a soft EoS. This results in
lower maximum masses and smaller radii. 

\begin{figure}
\begin{center}
\includegraphics[width=0.8\textwidth,angle=-90]{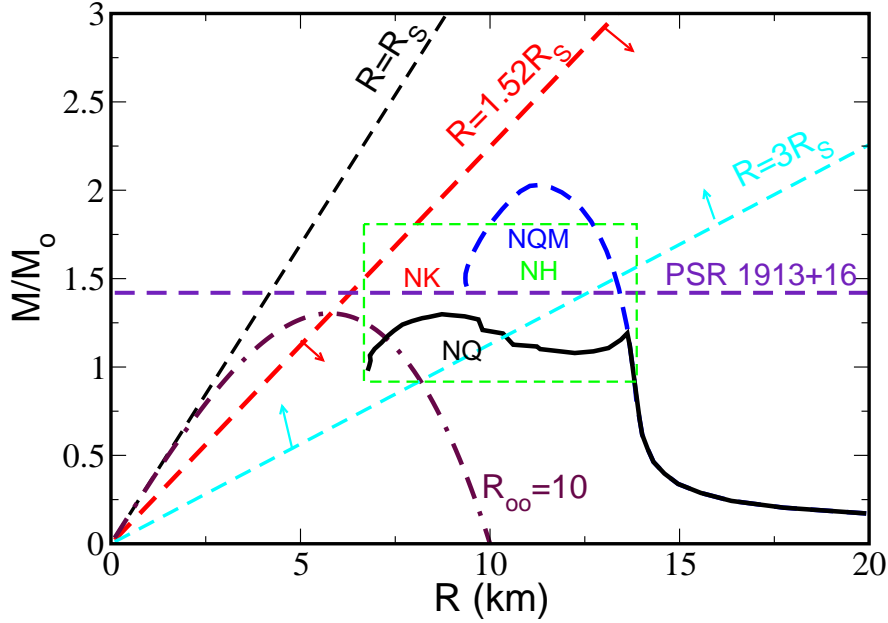}
\caption{Mass-radius relation for several exotic stars. The symbols
NK, NH and NQM refer to locations where we expect the maximum masses
of such stars to lie. They represent generic trends rather than
precise locations. The mean-field EoS is employed to represent the
nuclear part of the star.}
\label{masradiusexo}
\end{center} 
\end{figure}

This trend is shown Fig. \ref{masradiusexo}. The mass-radius
relationship for a hybrid compact object containing nuclear matter at
low density and pure quark matter at high density is shown (black
curve labeled NQ). Other exotic possibilities like hyperons (NH), kaon
condensation (NK) and phase transitions in which a mixed nuclear-quark
phase is favored (NQM) have a similar effect on the mass-radius
curve. Kaon condensation results in stars that are relatively compact while
the hyperonic and the nuclear-quark mixed phase stars favor modestly
larger maximum masses and radii. The rough estimate of the region of
the M-R diagram which can be populated by exotic stars given the
uncertainties in these models is also indicated by the dashed box.
Mass and radius are integral quantities and are sensitive to the EoS
over a wide range of densities. Therefore, in principle one requires a
large number of different and simultaneous mass-radius measurements to
directly infer the EoS. However, our knowledge of the nuclear
properties at nuclear saturation density and nuclei provide valuable
additional information. This is encoded in most realistic equations of
state and lead to the generic trends described above. The accurate
measurement of both the mass and radius of star will provide a very
stringent constraint on the high density EoS. For example, the
discovery of a compact object with a $R \lsim 8$ km would provide strong
support for the existence of novel phases inside neutron stars. On the
other hand, the accurate measurement of neutron star mass $M \gsim
1.8~M_{\odot}$ would support the absence of phase transitions.
While neither of these hypothetical measurements will provide
conclusive evidence, they will provide valuable guidance to the
theoretical models of high-density matter.

\section{Supernova Neutrinos} 

Neutrinos play an important role in stellar evolution. By virtue of
their weak interactions with matter, neutrinos provide a mechanism for
energy loss from dense stellar interiors. In neutron stars neutrinos
are responsible for most of the energy radiated from their birth in a
supernova explosion through several thousand years of subsequent
evolution. In this section we present an overview of some of the
nuclear/particle physics issues that play a role in understanding the
rate of propagation and production of neutrinos inside neutron
stars. The calculation of these rates is of current interest since
several research groups are embarking on large-scale numerical
simulations of supernova and neutron star evolution
\cite{Rampp:2002kn,Burrows:2002bz,Mezzacappa:2000jb}.  Even moderate
changes in the nuclear microphysics associated with the weak
interaction rates at high density can impact macroscopic features that
are observable. An understanding of the response of the strongly
interacting nuclear medium to neutrinos and its impact on neutron star
evolution promises to provide a means of probing the properties of the
dense medium itself \cite{Keil:hw,Pons:1998mm,Pons:2001ar}.

In this section we discuss neutrino interactions in dense matter
containing nucleons and leptons and neutrino interactions in exotic
new phases that are likely to occur in the dense inner core of the
neutron star. We emphasize the generic aspects of the
microphysics that affect the weak interaction rates and present
supporting qualitative arguments (For a detailed account see
Ref. \cite{Prakash:2000jr})
 
\subsection{Neutrino Interactions in Nucleonic Matter}
It was realized over a decade ago that the effects due to degeneracy
and strong interactions significantly alter the neutrino mean free
paths and neutrino emissivities in dense matter
\cite{sawy:75,frim:79,iwam:82,sawy:89}. However, it is only recently
that detailed calculations have become available
\cite{horo:91,raff:95,sigl:96,redd:97a,redd:97b,prak:97b,redd:98,redd:99a,burr:98,burr:99,hanh:00}.
The scattering and absorption reactions that contribute to the
neutrino opacity are
\begin{eqnarray}
\nu_e+B&\rightarrow& e^-+B' \,,\quad \quad
\bar{\nu}_e+B\rightarrow e^++B' \,, \nonumber \\
\nu_X+B&\rightarrow& \nu_X+B' \,,\quad \quad
\nu_X+e^-\rightarrow \nu_X +e^-  \,,
\label{nureact}
\end{eqnarray}
where the scattering reactions are common to all neutrino species and
the dominant source of opacity for the electron neutrinos is due to
the charged-current reaction. The important neutrino producing reactions in
the neutron star context are
\begin{eqnarray}
&e&^-+p\rightarrow n+\nu_e \,,\quad \quad \quad
n\rightarrow e^-+p+\bar{\nu}_e \,, \nonumber \\
&n&+n \rightarrow n+p+e^-+\bar{\nu}_e \,, \quad
n+n\rightarrow n+n+\nu_X+\bar{\nu_X} 
\label{chreact}
\end{eqnarray}

The weak interaction rates for the reactions in \Eqn{nureact} and
\Eqn{chreact} are modified in hot and dense matter because of many
in-medium effects.  The most important of these are: \\

\noindent (1) {\it Composition}: The rate for neutrino processes
depends sensitively on the composition of the medium, which is itself
sensitive to the nature of strong interactions.  First, the different
degeneracies of the different fermions determine the particle-hole
response due to Pauli blocking.  For example, a larger symmetry energy
favors higher proton fractions. This directly impacts the weak rates
because neutrinos couple differently to different baryonic species and
because the Pauli and momentum-conservation restrictions on rates
involving neutrons and protons are relaxed.  Consequently, the net
rates will depend on the individual concentrations. \\

\noindent (2) {\it In-medium dispersion relations}: At high density
the single-particle dispersion relations are significantly modified
from their non-interacting forms due to effects of strong interactions.
Interacting matter features smaller effective baryon masses and energy
shifts relative to non-interacting matter. This in turn affects the
weak interaction rates primarily because it modifies the density of
particle-hole states at the Fermi surface.\\

\noindent (3) {\it Correlations}: Low-energy neutrinos couple mainly
to the long wavelength density and spin-density fluctuations of the
strongly interacting nuclear plasma.  Repulsive particle-hole
interactions and Coulomb interactions generally result in a
suppression of the weak interaction rates since they increase the
energy cost associated with of such fluctuations. On the other hand,
interactions can also result in low-lying collective excitations to
which neutrinos can couple. This acts to increase the weak interaction
rates at low energy. Both of these effects may be calculated using the
Random Phase Approximation (RPA), in which particle-hole ring diagrams
are summed to all orders. Model calculations \cite{sawy:75,iwam:82,sawy:89,
horo:91,redd:97a,prak:97b,redd:99a,burr:98,burr:99} indicate that at
high density the neutrino cross sections are suppressed relative to
the case in which these effects are ignored.  In addition, these
correlations enhance the average energy transfer in neutrino-nucleon
collisions.  Improvements in determining many-body dynamic form
factors and assessing the role of particle-particle interactions in
dense matter at finite temperature are necessary before the full
effects of many-body correlations may be ascertained. \\

The relative importance of the various effects described above on
neutrino transport is only beginning to be studied systematically. As
a first step, we will focus on effects due to modifications (1)
through (3) above.  To see how this is accomplished we start with a
general expression for the differential cross section
~\cite{horo:91,redd:99a} for the processes shown in \Eqn{nureact}
\begin{eqnarray}
\frac {1}{V} \frac {d^3\sigma}{d^2\Omega_3 dE_3} &=& -\frac
{G_F^2}{128\pi^2} \frac{E_3}{E_1}~
\left[1-\exp{\left(\frac{-q_0-(\mu_2-\mu_4)}{k_BT}\right)}\right]^{-1}~
\nonumber \\ &\times & (1-f_3(E_3))~{\rm
Im}~(L^{\alpha\beta}\Pi^R_{\alpha\beta}) \,,
\label{dcross}
\end{eqnarray}
where the incoming neutrino energy is $E_{1}$, the outgoing lepton
energy is $E_{3}$ and the energy transfer $q_0=E_3-E_1$. The factor
$[1-\exp((-q_0-\mu_2+\mu_4)/k_BT)]^{-1}$ maintains detailed balance for
particle which are in thermal equilibrium at temperature $T$ and in
chemical equilibrium with chemical potentials $\mu_2$ and $\mu_4$,
respectively. The final state blocking of the outgoing lepton is
accounted for by the Pauli blocking factor $(1-f_3(E_3))$, where $f_3$
is the Fermi distribution function. The lepton tensor
$L_{\alpha\beta}$ is given by
\begin{equation}
L^{\alpha\beta}= 8[2k^{\alpha}k^{\beta}+(k\cdot q)g^{\alpha\beta}
-(k^{\alpha}q^{\beta}+q^{\alpha}k^{\beta})\mp i\epsilon^{\alpha\beta\mu\nu}
k^{\mu}q^{\nu}]\,,
\end{equation}
where $k_{\mu}$ is the incoming neutrino four-momentum and $q_{\mu}$
is the four-momentum transfer. In writing the lepton tensor, we have
neglected the electron mass terms, since typical electron energies are
of the order of a few hundred MeV.

The target-particle retarded polarization tensor is
\begin{equation}
{\rm Im} \Pi^R_{\alpha\beta} =
\tanh{\left(\frac{q_0+(\mu_2-\mu_4)}{2T}\right)} {\rm Im}~\Pi_{\alpha\beta}
\,,\\
\end{equation}
where $\Pi_{\alpha\beta}$ is the time-ordered (or causal) polarization
and is given by
\begin{equation}
\Pi_{\alpha\beta}(q)=-i \int \frac{d^4p}{(2\pi)^4} {\rm
Tr}~[T(S_2(p)\Gamma_{\alpha} S_4(p+q)\Gamma_{\beta})]\,.
\end{equation}
The Greens' functions $S_i(p)$ (the index $i$ labels particle species)
describe the propagation of baryons at finite density and temperature
\cite{horo:91}. The current operator $\Gamma_{\mu}$ is $\gamma_{\mu}$
for the vector current and $\gamma_{\mu}\gamma_5$ for the axial
current. Effects of strong and electromagnetic correlations may be
included by calculating the RPA polarization tensor
\begin{equation}
\Pi^{RPA} = \Pi + \Pi^{RPA} D \Pi~, 
\end{equation} 
where $D$ denotes the strong interaction matrix (see
Ref.~\cite{redd:99a} for more details).

\begin{figure}
\begin{center}
\includegraphics[width=0.9\textwidth,height=0.8\textwidth]{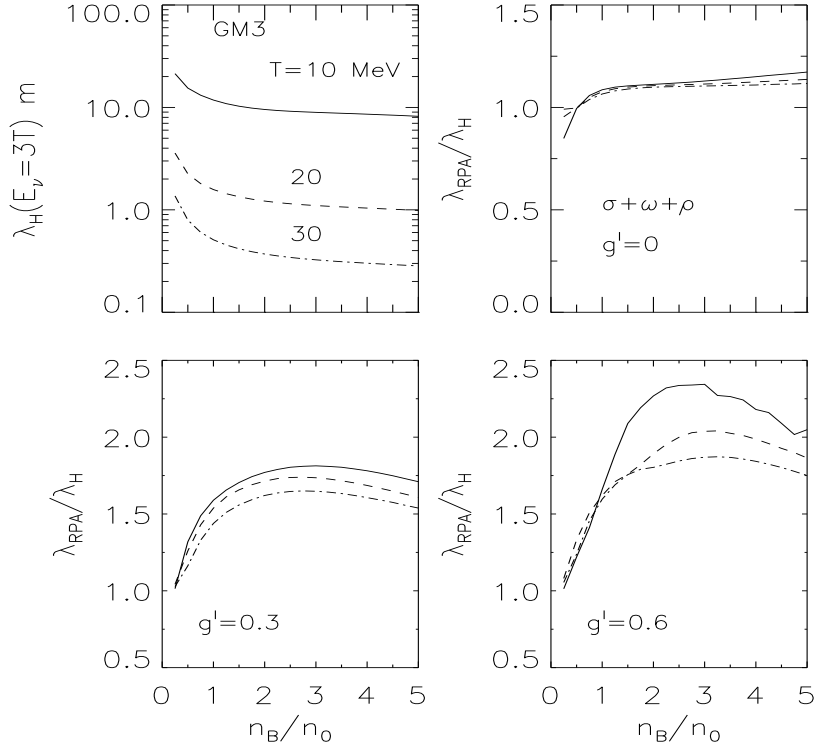}
\caption{Neutrino mean free paths in dense stellar matter containing
neutron, protons and electrons calculated in the relativistic RPA
($\lambda_{RPA}$) and normalized to those computed in the mean field or
Hartree approximation ($\lambda_H$) \cite{redd:99a}}.
\end{center}
\label{rpa}      
\end{figure}

Neutrino mean free path, which is inversely related to the cross
section per unit volume, calculated in relativistic RPA is shown in
Fig. \ref{rpa}. The model employed incorporates interactions via
$\sigma$, $\omega$, and $\rho$ exchange as in Walecka model
\cite{horo:91,redd:99a}.  It is supplemented by pion exchange and a
repulsive contact term, whose strength is parameterized by a constant
$g' \simeq 0.6$ to account for short-range spin-isospin
correlations. The results indicate that RPA corrections are most
significant in the spin-isospin channel and that low-temperature
correlation can suppress the cross section by a factor of two or
more. This is because, in the Fermi-gas limit, the axial-vector
response of medium makes a contribution that is roughly three times
larger than the vector response \cite{iwam:82}.  Quantitative
aspects of the suppression depend on the details of the model
employed. Nonetheless, we note that most model studies thus far
indicate similar suppression factors.
\subsection{Neutrino Interactions in Novel Phases}
In this section we explore how phase transitions impact the weak
interaction rates.  As discussed in earlier sections, novel phases of
baryonic matter are expected to occur at densities accessible in
neutron stars. These new phases include pion condensation, kaon
condensation, hyperons and quark matter. An understanding of how these
phases might influence neutrino propagation and emission is necessary
if we are to inquire if these phase transitions even occur inside
neutron stars. To illustrate how they neutrino rates we consider three
specific examples of phase transitions: (1) generic first-order
transitions; (2) superconducting quark matter and (3)
color-flavor-locked superconducting quark matter.

\subsubsection{Inhomogeneous Phases: Effects of First-Order Transitions}
First-order phase transitions in neutron stars can result in the
formation of heterogeneous phases in which a positively-charged
nuclear phase coexists with a negatively-charged new phase which is
favored at higher densities \cite{Glendenning:1992vb}. This is a
generic feature of first-order transitions in matter with two
conserved charges. In the neutron star context these correspond to
baryon number and electric-charge. Reddy, Bertsch and Prakash
\cite{RBP} have studied the effects of inhomogeneous phases on
$\nu$-matter interactions. Based on simple estimates of the surface
tension between nuclear matter and the exotic phase, typical droplet
sizes range from $5-15$ fm \cite{size}, and interdroplet spacings are
several times larger. The propagation of neutrinos whose wavelength
is greater than the typical droplet size and less than the
interdroplet spacing, i.e., $ 2{\rm~MeV}\le E_\nu \le 40{\rm~MeV} $,
will be greatly affected by the heterogeneity of the mixed phase, as a
consequence of the coherent scattering of neutrinos from the droplet.

The Lagrangian that describes the neutral-current coupling of
neutrinos to the droplet is
\begin{equation}
{\mathcal{L}}_W = \frac{G_F}{2\sqrt{2}} ~\bar{\nu}\gamma_\mu(1-\gamma_5) \nu
~J^{\mu}_D \,,
\label{nuD}
\end{equation}
where $J^{\mu}_D$ is the neutral weak current carried by the droplet
and $G_F=1.166 \times 10^{-5}$ GeV$^{-2}$ is the Fermi weak coupling
constant. For non-relativistic droplets, $J^{\mu}_D$ has only a
time-like component, $\rho_W(x)$, where $\rho_W$ is the excess weak
charge density in the droplet. The total weak charge enclosed in a
droplet of radius $r_d$ is $N_W=\int_0^{r_d} d^3x ~\rho_W(x)$ and the
associated form factor is
\begin{equation}
F(q)=(1/N_W)\int_0^{r_d} d^3x ~\rho_W(x)~ \sin{qx}/qx \,.
\end{equation}
The differential cross section for neutrinos scattering from an
isolated droplet is then
\begin{equation}
\frac{d\sigma}{d\cos{\theta}}= \frac{E_\nu^2}{16\pi}
G_F^2 N^2_W(1+\cos{\theta}) F^2(q) \,.
\label{diff}
\end{equation}
In the above equation, $E_\nu$ is the neutrino energy and $\theta$ is the
scattering angle. Since the droplets are massive, we consider only elastic
scattering for which the magnitude of the three-momentum transfer is 
$q=\sqrt{2}E_\nu(1-\cos{\theta})$.

We must embed the droplets in the medium to evaluate the neutrino
transport parameters.  The droplet radius $r_d$ and the inter-droplet
spacing are determined by the interplay of surface and Coulomb
energies.  In the Wigner-Seitz approximation, the unit-cell radius is
$R_W=(3/4\pi N_D)^{1/3}$ where the droplet density is $N_D$. 
Multiple droplet scattering cannot be neglected for
$E_\nu \lsim 1/R_W$. The effects of other droplets is to cancel
scattering in the forward direction, because the interference is
destructive except at exactly zero degrees, where it only produces a change
in the index of refraction of the medium. 

These effects are usually incorporated by multiplying the differential
cross section Eq.~(\ref{diff}) by the static form factor of the
medium. The static form factor, defined in terms of the radial
distribution function of the droplets, $g(r)$, is
\begin{equation}
S(q)= 1 + N_D \int d^3r \exp{i \vec{q}.\vec{r}}~[ g(r)-1] \,.
\end{equation}
The droplet correlations, which determine $g(r)$, arise due to the
Coulomb force and are measured in terms of the dimensionless Coulomb
number $\Gamma=Z^2e^2/(8\pi R_W kT)$. Due to the long-range character
of the Coulomb force, the role of screening, and the finite droplet
size, $g(r)$ cannot be computed analytically. We use a simple form for
the radial distribution function $g(r) = \theta(r-R_W)$. This choice
for $g(r)$ is equivalent to subtracting, from the weak charge density
$\rho_W$, a uniform density which has the same total weak charge $N_W$
as the matter in the Wigner-Seitz cell. Thus, effects due to $S(q)$
may be incorporated by replacing the form factor $F(q)$ by
\begin{eqnarray}
\tilde{F}(q) = F(q) - 3~
\frac{\sin{qR_{W}} - (qR_{W})\cos{qR_{W}}}{(q R_{W})^3}  \,.
\label{formc}
\end{eqnarray}
The neutrino--droplet differential cross section per unit volume
then follows:
\begin{equation}
\frac{1}{V}\frac{d\sigma}{d\cos{\theta}}=
N_D~\frac{E_\nu^2}{16\pi} G_F^2 N^2_W(1+\cos{\theta}) \tilde{F}^2(q) \,.
\label{diff1}
\end{equation}
Note that even for small droplet density $N_D$, the factor $N_W^2$
acts to enhance the droplet scattering since, as we shall see below $N_W
\sim 800$.

To quantify the importance of droplets as a source of opacity, we
compare with the standard scenario in which matter is uniform and
composed of neutrons. The dominant source of opacity is then due to
scattering from thermal fluctuations and, in the non-relativistic limit 
\Eqn{dcross} reduces to 
\begin{eqnarray}
\frac {1}{V}\frac{d\sigma}{d\cos{\theta}} &=&
\frac{3G_F^2~E_{\nu}^2}{16\pi}\left[c_V^2(1+\cos{\theta})
+(3-\cos{\theta})c_A^2\right]~n_n~\left[\frac{k_BT}{E_{Fn}}\right]\,,
\label{diff2}
\end{eqnarray}
where $c_V$ and $c_A$ are respectively the vector and axial coupling
constants of the neutron, $n_n$ is the neutron number density,
$E_{Fn}=k_{Fn}^2/2M_n$ is the neutron Fermi energy and $T$ is the
matter temperature \cite{iwam:82}. The transport cross sections that
are employed in studying the diffusive transport of neutrinos in the
core of a neutron star are differential cross sections weighted by the
angular factor $(1-\cos{\theta})$. The transport mean free path,
$\lambda(E_\nu)$, for a given neutrino energy $E_\nu$ is given by
\begin{eqnarray}
\frac{1}{\lambda(E_\nu)}=\frac{\sigma_T(E_\nu)}{V} =
\int d\cos{\theta}~ (1-\cos{\theta})
\left[\frac{1}{V} \frac{d\sigma}{d\cos{\theta}}\right] \,.
\label{fint}
\end{eqnarray}

\begin{figure}
\resizebox{0.49\textwidth}{!}{%
\includegraphics{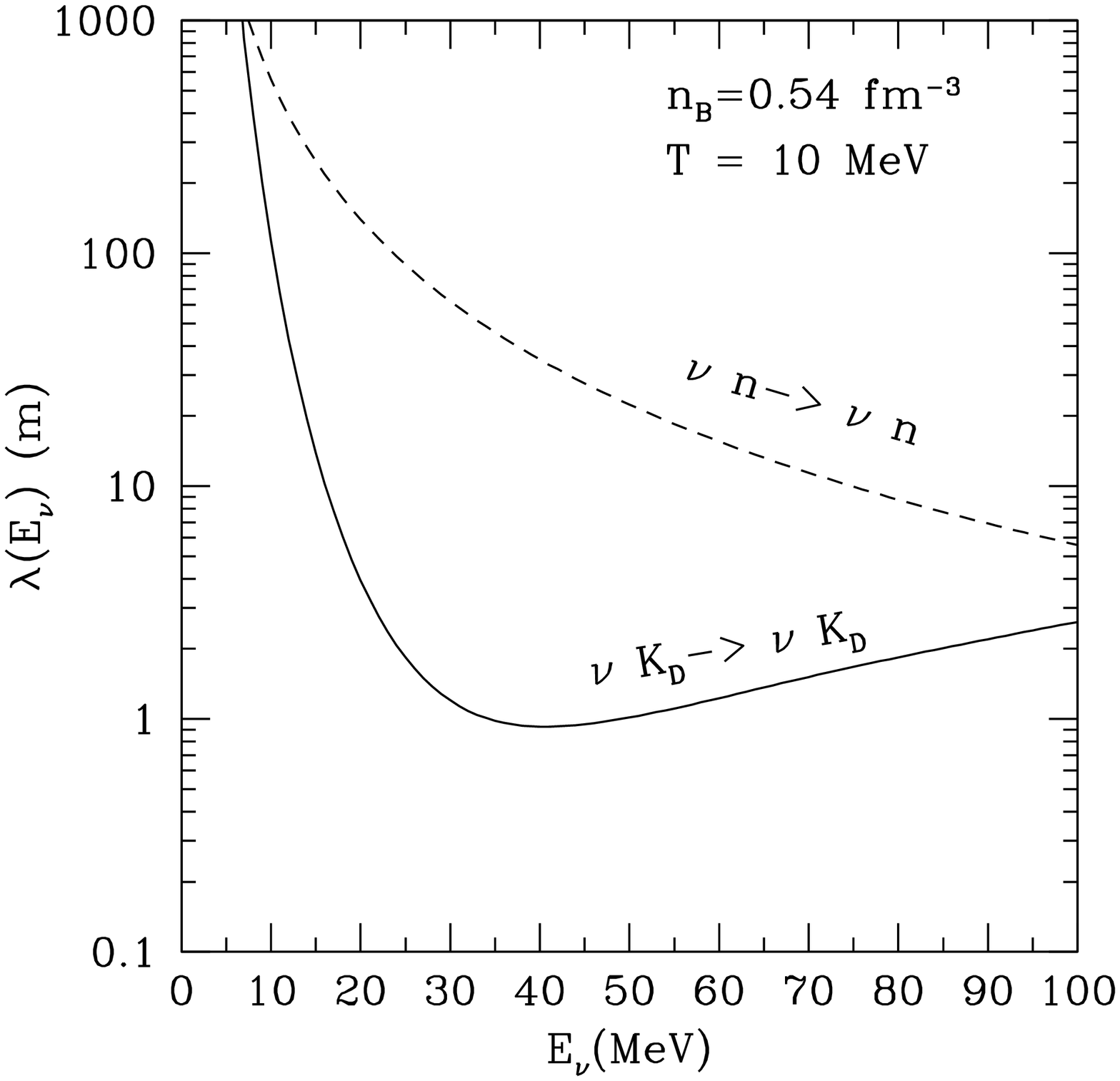}
}
\resizebox{0.46\textwidth}{!}{%
\includegraphics{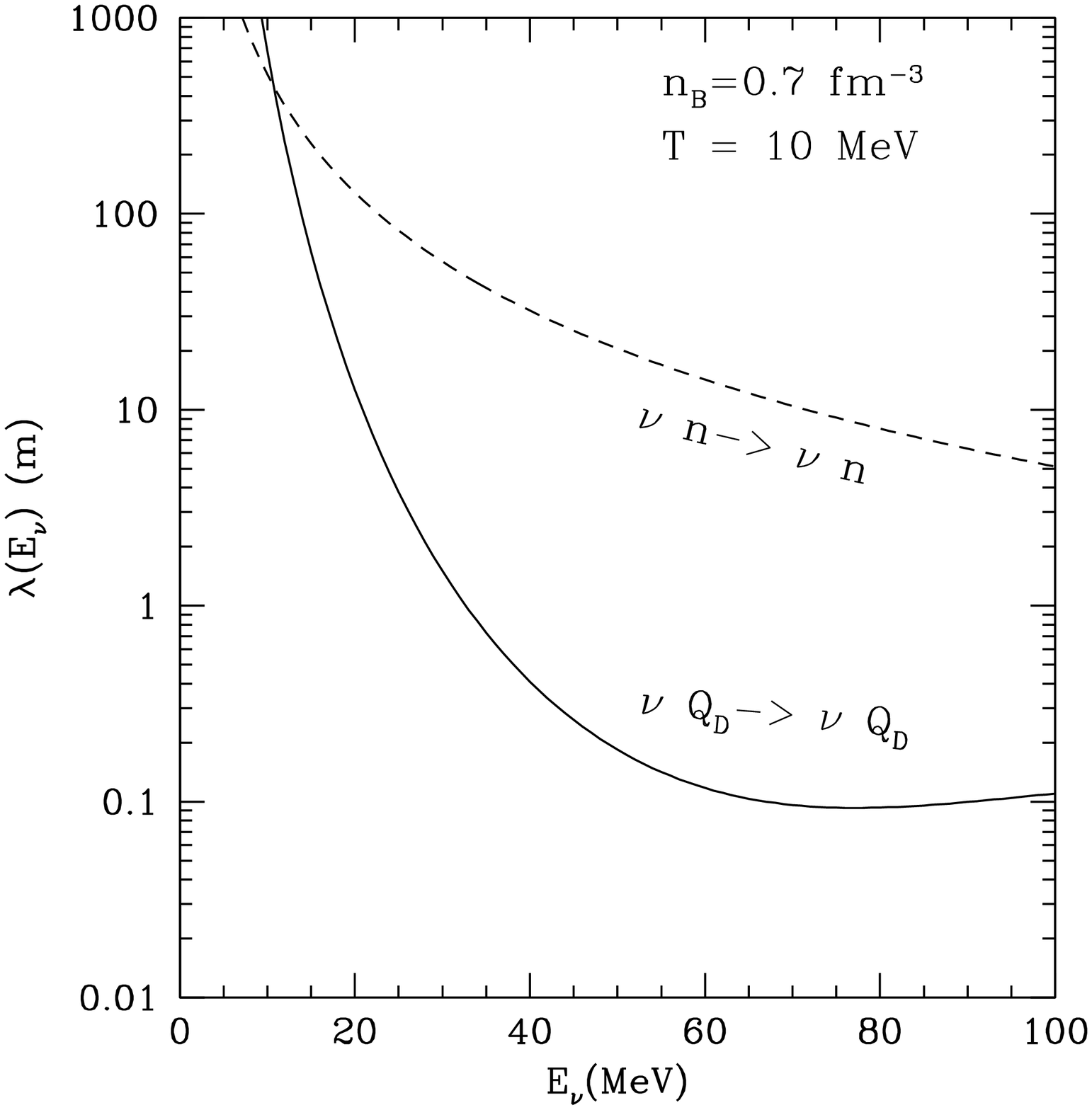}
}
\caption{Neutrino mean free paths as a function of neutrino energy.
Solid lines are for matter in a mixed phase containing kaons (left
panel) and quarks (right panel), and dashed curves are for uniform matter.}
\label{mpaths}
\end{figure}

Models of first-order phase transitions in dense matter provide the
weak charge and form factors of the droplets and permit the evaluation
of $\nu$--droplet scattering contributions to the opacity of the mixed
phase. We consider two models, namely the first-order kaon condensate
and the quark-hadron phase transition. The neutrino mean free paths in
the mixed phase are shown in the left and right panels of
Fig.~\ref{mpaths}, respectively.  The results are shown for the
indicated values of the baryon density $n_B$ and temperature $T$,
where the model predicts a mixed phase exists. The kaon droplets are
characterized by radii $r_d\approx 7$ fm and inter-droplet spacings
$R_W\approx 20$ fm, and enclose a net weak charge $N_W \approx
700$. The quark droplets are characterized by $r_d\approx 5$ fm, and
$R_W\approx 11$ fm, and an enclosed excess weak charge $N_W \approx
850$.  For comparison, the neutrino mean free path in uniform neutron
matter at the same $n_B$ and $T$ are also shown. The typical neutrino
energy for neutrino in thermal equilibrium is given by $E_\nu \approx
\pi T$ (the factor $\pi$ here is just a convenient approximation). For
these energies, it is apparent that there is a large coherent
scattering-induced reduction in the mean free path. At much lower
energies, the interdroplet correlations tend to screen the weak charge
of the droplet, and at higher energies the coherence is attenuated by
the droplet form factor.

The large reduction in neutrino mean free path found here implies that
a mixed phase will cool significantly slower than homogeneous
matter. Consequently, the observable neutrino luminosity at late times
might be affected, as it is driven by the transport of energy from the
deep interior.

\subsubsection{Effects of Quark Superconductivity}
As discussed previously, recent theoretical works
\cite{Barrois:1977xd,BigGap} suggest that quarks form Cooper pairs in
medium, a natural consequence of attractive interactions destabilizing
the Fermi surface. Model calculations, mostly based on four-quark
effective interactions (Nambu-Jona-Lasinio--like models), predict the
restoration of spontaneously broken chiral symmetry through the onset
of color superconductivity at relatively low temperatures.  They
predict an energy gap of $\Delta \sim 100$ MeV for a typical quark
chemical potential of $\mu \sim 400 $ MeV.  As in BCS theory, the
gap will weaken for $T > 0$, and at some critical temperature $T_c$
there is a (second-order) transition to a normal or unpaired
quark-gluon plasma.  During cooling from an initial temperature in
excess of $T_c$, the formation of a gap in the fermion excitation
spectrum in quark matter will influence various transport properties
of the system. Carter and I have studied its influence on the
transport of neutrinos \cite{CR00}.

The differential neutrino scattering cross section per unit volume in
an infinite and homogeneous system of relativistic fermions as
calculated in linear response theory is given by \Eqn{dcross}.
The medium is characterized by the quark polarization tensor
$\Pi_{\alpha\beta}$.  In the case of free quarks, each flavor
contributes a term of the form
\begin{equation}
\Pi_{\alpha\beta}(q)=-i {\rm Tr}_c \int
\frac{d^4p}{(2\pi)^4} {\rm Tr}~[S_0(p)\Gamma_{\alpha} 
S_0(p+q)\Gamma_{\beta}] \,, 
\label{pi_free}
\end{equation}
where $S_0(p)$ is the free quark propagator. At finite chemical
potential and temperature the quark propagator is given by
\begin{equation}
S_0(p)^{b g}_{a f}= i\delta_a^b \delta^g_f~
\left(\frac{\Lambda^+(p)}{p_0^2 - E_p^2} + 
\frac{\Lambda^-(p)}{p_0^2 - \bar{E}_p^2}\right)
~(p_\mu\gamma^\mu -\mu \gamma_0)\,.
\label{s_0}
\end{equation}
This is written in terms of the particle and anti-particle projection
operators $\Lambda^+(p)$ and $\Lambda^-(p)$ respectively, where
$\Lambda^{\pm}(p)=(1 \pm \gamma_0\vec{\gamma} \cdot \hat{p})/2$.  The
excitation energies are simply $E_p = |\vec{p}|-\mu$ for massless
quarks and $\bar{E}_p = |\vec{p}|+\mu$ for massless anti-quarks. The
outer trace in \Eqn{pi_free} is over color and simplifies to $3$.
The inner trace is over spin, and the $\Gamma_\alpha$ are the
neutrino-quark vertex functions.  Specifically, the vector
polarization is computed by choosing $(\Gamma_{\alpha},
\Gamma_{\beta}) = ( \gamma_{\alpha}, \gamma_{\beta} )$.  The axial and
mixed vector-axial polarizations are similarly obtained from
$(\Gamma_{\alpha}, \Gamma_{\beta}) = (\gamma_{\alpha}\gamma_5,
\gamma_{\beta}\gamma_5)$ and $(\Gamma_{\alpha}, \Gamma_{\beta}) =
(\gamma_{\alpha}, \gamma_{\beta}\gamma_5)$, respectively.

The free quark propagators in \Eqn{pi_free} are modified in a
superconducting medium. In calculating these effects, we will consider
the simplified case of QCD with two quark flavors which obey $SU(2)_L
\otimes SU(2)_R$ flavor symmetry, given that the light $u$ and $d$
quarks dominate low-energy phenomena.  Furthermore we will assume
that, through some unspecified effective interactions, quarks pair in
a manner analogous to the BCS mechanism \cite{Bardeen:1957mv}.  The
relevant consequences of this are the restoration of chiral symmetry
(hence all quarks are approximately massless) and the existence of an
energy gap at zero temperature, $\Delta_0$. As discussed earlier in
section \ref{superquark}, this superconducting phase is called the 2SC
phase. The approximate temperature dependence of the gap is taken from
BCS theory
\begin{equation}
\Delta(T) = \Delta_0 \sqrt{ 1 - \left(\frac{T}{T_c}\right)^2 }\,,
\label{deltabcs}
\end{equation}
where the critical temperature $T_c \simeq 0.57 \Delta_0$; this
relation has been shown to hold for perturbative QCD and is thus a
reasonable assumption for non-perturbative physics.  Since the scalar
diquark (in the $\bar{\bf 3}$ color representation) is the most
attractive channel, we consider the anomalous propagator
\cite{propagators}
\begin{eqnarray}
F(p)_{a b f g} &=&  
\langle q_{f a}^T(p) C\gamma_5 q_{g b}(-p) \rangle \nonumber\\
&=& -i \epsilon_{a b 3} \epsilon_{fg} 
\Delta \left(\frac{\Lambda^+(p)}{p_0^2 - \xi_p^2} + 
\frac{\Lambda^-(p)}{p_0^2 - \bar{\xi}_p^2}\right) \gamma_5~C\,. 
\label{a_bcs}
\end{eqnarray}
Here, $a,b$ are color indices, $f,g$ are flavor indices,
$\epsilon_{abc}$ is the usual anti-symmetric tensor and we have
conventionally chosen 3 to be the condensate color.  The
quasi-particle energy is $\xi_p = \sqrt{(|\vec{p}|-\mu)^2 +
\Delta^2}$, and for the anti-particle $\bar\xi_p =
\sqrt{(|\vec{p}|+\mu)^2 + \Delta^2}$. This propagator is also
antisymmetric in flavor and spin, with $C=-i\gamma_0\gamma_2$ being
the charge conjugation operator. 

The color bias of the condensate forces a splitting of the normal
quark propagator into colors transverse and parallel to the diquark.
Quarks of color 3, parallel to the condensate in color space, will be
unaffected and propagate freely. Their response is characterized by
the polarization tensor given in \Eqn{pi_free}, but without the
trace over color space since it involves only quarks of one color.  On
the other hand, transverse quark colors 1 and 2 participate in the
diquark and thus their quasi-particle propagators are given as
\begin{equation}
S(p)^{b g}_{a f}= i\delta_a^b \delta^g_f~
\left(\frac{\Lambda^+(p)}{p_o^2 - \xi_p^2} + 
\frac{\Lambda^-(p)}{p_o^2 - \bar{\xi}_p^2}\right)
~(p_\mu\gamma^\mu -\mu \gamma_0)\,.
\label{s_bcs}
\end{equation}

The appearance of an anomalous propagator in the superconducting phase
indicates that the polarization tensor gets contributions from both
the normal quasi-particle propagators Eq.~(\ref {s_bcs}) and anomalous
propagator Eq.~(\ref{a_bcs}).  Thus, to order $G_F^2$,
Eq.~(\ref{pi_free}) is replaced with the two contributions
corresponding to the normal and anomalous diagrams and is given by
\begin{eqnarray}
\Pi_{\alpha\beta}(q) &=& -i \!\int\! \frac{d^4p}{(2\pi)^4} \left\{
{\rm Tr}~[S_0(p)\Gamma_{\alpha} S_0(p+q)\Gamma_{\beta}] 
\right. \nonumber \\ 
&+& \left.
2 {\rm Tr}~[S(p)\Gamma_{\alpha} S(p+q)\Gamma_{\beta}]
\right. \nonumber\\
&+& \left. 
2 {\rm Tr}~[F(p)\Gamma_{\alpha} \bar{F}(p+q)\Gamma_{\beta}]
\right\} \,.
\label{pi_bcs}
\end{eqnarray}
The $S(p)$-$F(p)$ mixed terms vanish and the remaining trace is over
spin, as the color trace has been performed giving the factors of 2. 

For neutrino scattering we must consider vector, axial, and mixed
vector-axial channels, all summed over flavors.  
The full polarization, to be used in evaluating Eq.~(\ref{dcross}),
may be written
\begin{equation}
\Pi_{\alpha\beta} = \sum_f\,\left[ (c_V^f)^2 \Pi^V_{\alpha\beta} +
(c_A^f)^2 \Pi^A_{\alpha\beta} - 2 c_V^f c_A^f\Pi^{VA}_{\alpha\beta}\right]\,.
\label{polsum}
\end{equation}
The coupling constants for up quarks are $c_V^u =
\textstyle{\frac{1}{2}} - {\textstyle{\frac{4}{3}}} \sin^2\theta_W$
and $c_A^u = \textstyle{\frac{1}{2}}$ , and for down quarks, $c_V^d =
-\textstyle{\frac{1}{2}}+\textstyle{\frac{2}{3}}\sin^2\theta_W$ and
$c_A^d = -\textstyle{\frac{1}{2}}$, where $\sin^2\theta_W = 0.23$
is the Weinberg angle.

\begin{figure}[t]
\begin{center}
\resizebox{0.8\textwidth} {!}{%
\includegraphics{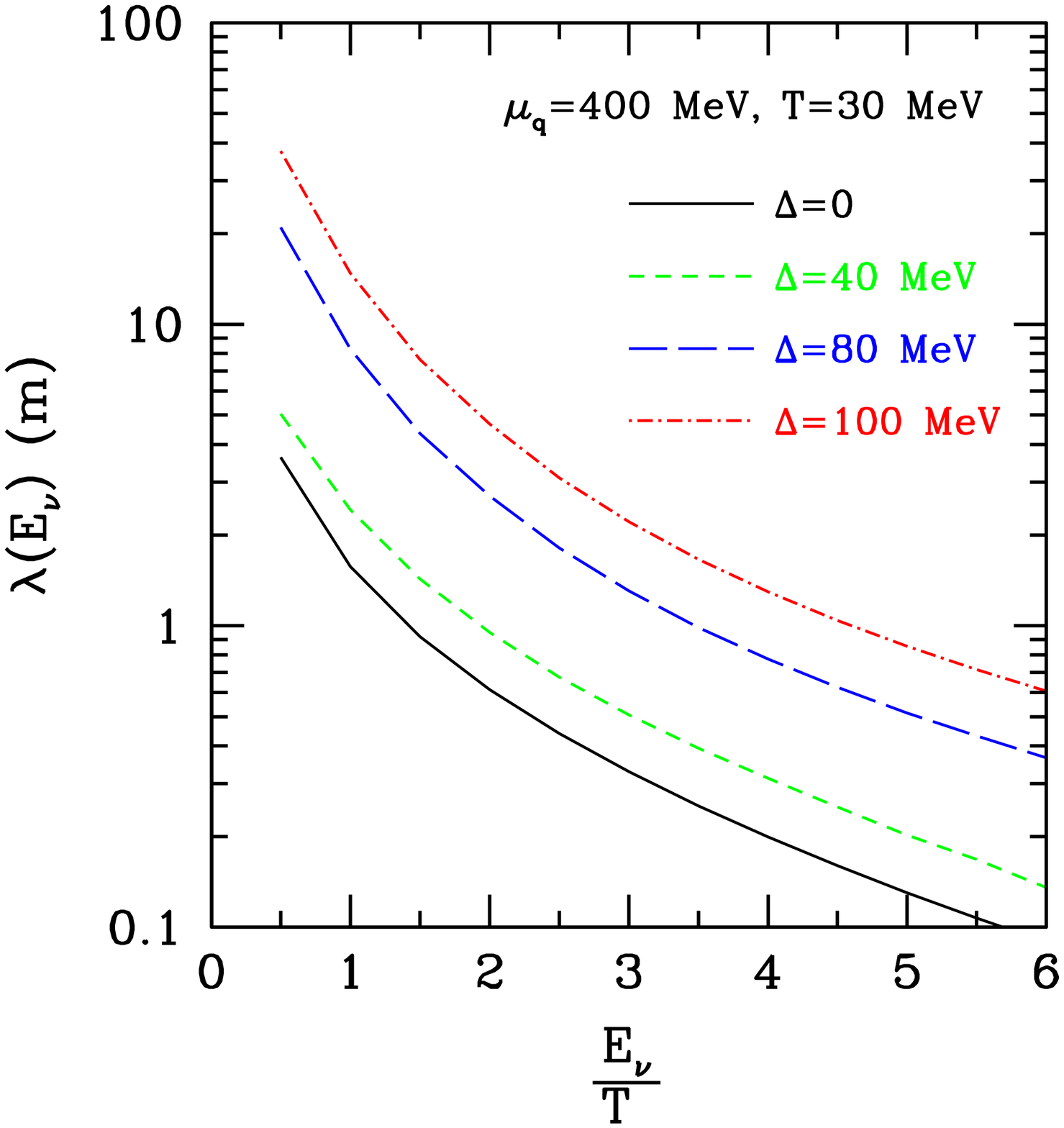}
\includegraphics{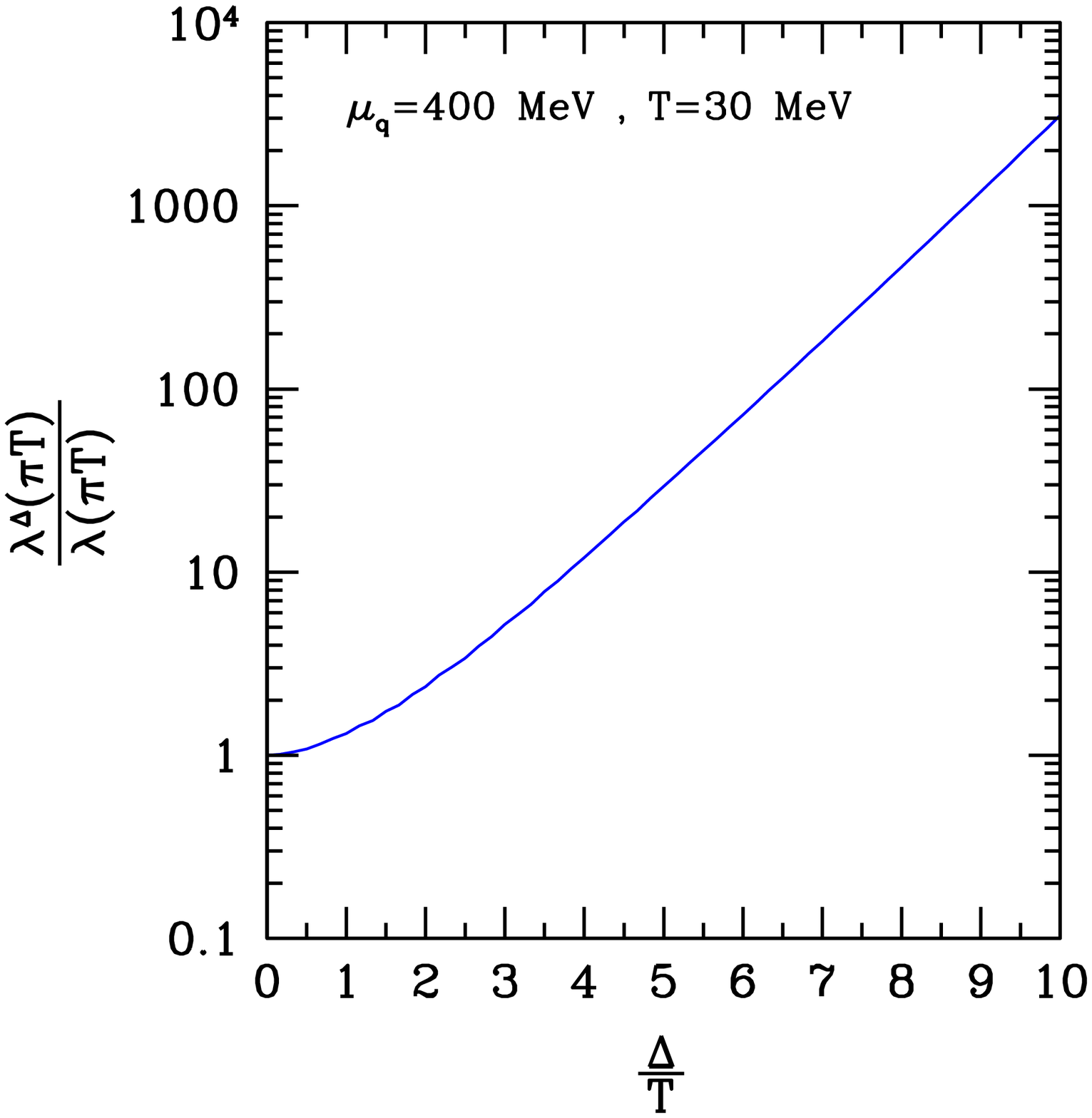}
}
\end{center}
\caption{Left panel: Neutrino mean free path as a function of neutrino
energy $E_\nu$.  Right panel: Neutrino mean free paths for $E_\nu=\pi
T$ as a function of $\Delta(T)/T$.  These results are virtually
independent of temperature for $T \lsim 50$ MeV. $\lambda^{\Delta}$
and $\Lambda$ denote the mean free paths in the superconducting and
normal phases respectively.}
\label{lambda}
\end{figure}

The differential cross section, Eq.~(\ref{dcross}), and the total
cross section are obtained by integrating over all neutrino energy
transfers and/or angles.  Results for the neutrino mean free path,
$\lambda=V/\sigma$, are shown in Fig.~\ref{lambda} as a function of
incoming neutrino energy $E_\nu$ (for ambient conditions of $\mu=400$
MeV and $T=30$ MeV).  They show the same energy dependence found
previously for free relativistic and degenerate fermion matter
\cite{redd:98}; $\lambda\propto1/E_\nu^2$ for $E_\nu \gg T$ and
$\lambda\propto1/E_\nu$ for $E_\nu \ll T$.  The results indicate that
this energy dependence is not modified by the presence of a gap when
$\Delta \sim T$.  Thus, the primary effect of the superconducting
phase is a much larger mean free path.  This is consistent with the
suppression found in the vector-longitudinal response function,
$\Pi^V_{00}$, which dominates the polarization sum Eq.~(\ref{polsum}),
at $q_0<q$.
\subsubsection{Neutrino Interactions with Collective Excitations (Goldstone modes)}
The discussion in the preceding section assumed that there were no low
energy collective excitations to which the neutrinos could couple.  As
discussed in Sec.~\ref{superquark}, this is true in the two-flavor
superconducting (2SC) phase of quark matter. For three flavors and
when the strange quark mass is negligible compared to the chemical
potential, the ground state is characterized by pairing that involves
all nine quarks in a pattern that locks flavor and color
\cite{Alford:1998mk}. Naively we can expect significant differences in
the weak interaction rates between the normal and the CFL phases of
quark matter since the latter is characterized by a large gap in the
quark excitation spectrum.  However, as detailed in
Sec. \ref{superquark}, diquark condensation in the CFL phase breaks both
baryon number and chiral symmetries. The Goldstone bosons that arise
as a consequence introduce low-lying collective excitations to the
otherwise rigid state.  Thus, unlike in the normal phase where quark
excitations near the Fermi surface provide the dominant contribution
to the weak interaction rates, in the CFL phase it is the dynamics of
the low-energy collective states-- the Goldstone bosons --that are
relevant. Neutrino interactions with Goldstone modes have been
recently investigated by Jaikumar et al. \cite{Jaikumar:2002vg} and
Reddy et al. \cite{Reddy:2002xc}.

There are several articles that describe in detail the effective
theory for Goldstone modes in CFL quark matter \cite{CFLmesons}.  We
will not review them here except to note that it is possible to
parameterize low energy excitations about the $SU(3)$ symmetric CFL
ground state in terms of the two fields $B=H / \sqrt{24}f_H$ and
$\Sigma=\exp{2i({\lambda \cdot \pi}/f_\pi+\eta'/f_A)}$. The $B$ fields
represent Goldstone modes of broken baryon number $H$. The $\Sigma$
field of broken chiral symmetry, i.e, the pseudo-scalar octet of Goldstone
modes $\pi$, and the pseudo-Goldstone boson $\eta'$.

The massless Goldstone boson associated with spontaneous breaking of
$U(1)_B$ couples to the weak neutral current. This is because the weak
isospin current contains a flavor singlet component. Although
neutrinos couple to the flavor octet of Goldstone bosons, it turns out
that the neutrino mean free path is mostly determined by processes
involving the massless baryon number Goldstone mode
\cite{Reddy:2002xc}. For this reason, we focus our attention on
these latter processes.  The amplitude for processes involving the
$U(1)_B$ Goldstone boson $H$ and the neutrino neutral current is given
by \cite{Reddy:2002xc}
\begin{equation} 
A_{H \nu \bar{\nu}}=
~\frac{4}{\sqrt{3}}~G_F~f_H ~ \tilde{p}_{\mu}~j_Z^{\mu}\,, 
\end{equation}

\noindent where $\tilde{p}_{\mu}=(E,v^2 \vec{p})$ is the modified four
momentum of the Goldstone boson and $v=c/\sqrt{3}$ is its velocity
\cite{Reddy:2002xc}. The decay constants for the $U(1)_B$ and the
pseudo-scalar octet of Goldstone modes have also been computed in
earlier work \cite{CFLmesons} and are given by $f^2_H=3 \mu^2/8 \pi^2$
and $f^2_{\pi}=(21-8 \ln 2)\mu^2/36 \pi^2$, respectively.

The neutrino mean free path due to the reaction $\nu \rightarrow H \nu$
can be calculated analytically and is given by
\begin{equation} 
\frac{1}{\lambda_{\nu\rightarrow H \nu}(E_\nu)}=
~\frac{256}{45\pi}~\left[\frac{v(1-v)^2(4+v)}{4~(1+v)^2}
\right]~G_F^2~f_H^2
~~E_{\nu}^3 
\end{equation}

Neutrinos of all energies can absorb a thermal mode and scatter into
either a final state neutrino by neutral current processes like $\nu +
H \rightarrow \nu$ and $\nu+ \pi^0 \rightarrow \nu$ or via the
charged-current reaction into a final state electron by the process
$\nu_e + \pi^- \rightarrow e^-$.  These processes are temperature
dependent, since they are proportional to the density of Goldstone
modes in the initial state. The mean free path due to these processes,
which we collectively refer to as Cerenkov absorption, can be
computed.  Reactions involving the $H$ boson dominate over other
Cerenkov absorption processes due to their larger population and
stronger coupling to the neutral current \cite{Reddy:2002xc}. For this
case we find the neutrino mean free path is given by
\begin{eqnarray}
\frac{1}{\lambda_{\nu H \rightarrow \nu}(E_\nu)}&=&
F(v,\gamma)G_F^2~f_H^2~E_\nu^3 \nonumber \\
F(v,\gamma)&=& 
\frac{128}{3\pi}\left[\frac{v~(1+v)^2}{(1-v)}\right] 
\left[g_2(\gamma)+
\frac{2v}{(1-v)}g_3(\gamma)-\frac{(1+v)}{(1-v)}g_4(\gamma)\right]
\end{eqnarray} 
where $\gamma=2vE_\nu/(1-v)k_BT$ and the integrals $g_n(\gamma)$ 
are defined by 
\begin{equation}
g_n(\gamma) = \int_{0}^{1} dx ~\frac{x^n}{\exp{(\gamma x)}-1} ~\,. 
\end{equation}

In contrast to processes involving the emission or absorption of the
Goldstone modes by neutrinos, the usual scattering process involves
the coupling of the neutrino current to two mesons. The amplitude for
these processes vanishes for the $H$ meson and is suppressed by the
factor $p/f_\pi$ where $p$ is the mode momentum for the flavor octet
Goldstone modes.  Fig.~14 shows the contribution of all
Goldstone boson-neutrino processes contributing to the neutrino mean
free path in the CFL phase, including the dominant contribution arising
from processes involving the massless mode.

\begin{figure}[t]
\begin{center}
\resizebox{0.8\textwidth}{!}{
\includegraphics[clip]{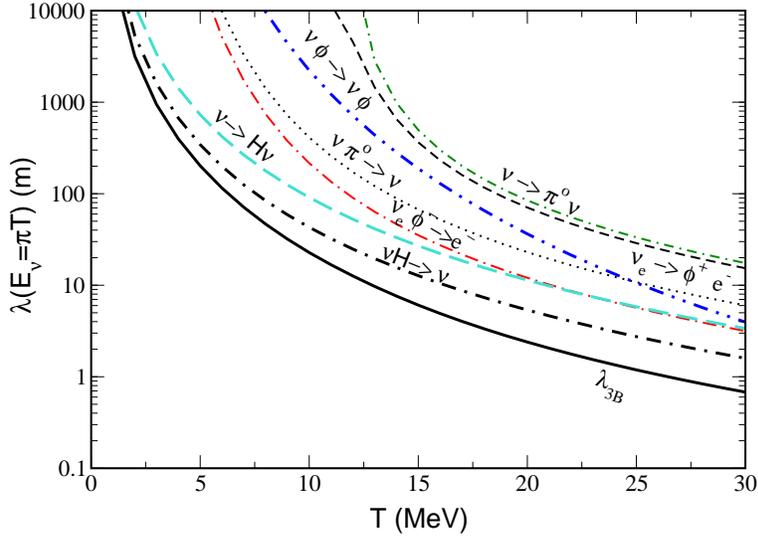}
}
\caption{Neutrino mean free path in a CFL meson plasma as a function
of temperature. The neutrino energy $E_\nu=\pi T$ and is
characteristic of a thermal neutrino. $\lambda_{3B}$ is the total
contribution of Cerenkov (absorption and emission) processes to the
neutrino mean free path \cite{Reddy:2002xc} and $\phi$ refers
collectively to the octet of Goldstone modes.}
\end{center}
\label{path}
\end{figure}

It is interesting to note that the existence of one massless Goldstone mode
compensates for the large gap in the particle-hole excitation
spectrum. The contrast between the findings of the previous section
where no low energy Goldstone modes coupled to neutrinos to those
presented here is striking. The mean free path in the CFL phase is
surprisingly similar to that in normal non-superconducting quark
phase.

\section{Conclusions}
Neutron stars are excellent laboratories to study possible phase
transitions at high baryon density. We have attempted to provide a
glimpse of the rich phenomena that may arise if such transitions were
to occur inside these compact objects. The real challenge lies in
being able to identify those characteristics of the phase transitions
that will uniquely affect observable aspects of the neutron star
structure and evolution. The structural properties of the star depend
on the EoS of high-density matter. Generically, we found that phase
transitions result in softening of the high-density EoS. In this case,
the maximum mass and radius of neutron stars are lowered. 

Neutrino production and propagation in dense matter directly impact
neutron stars thermal evolution. The presence of novel phases at high
density influences the low-energy weak interaction rates. For specific
examples considered in this article we found that the rates could be
either greatly enhanced or reduced. In some instances, the neutrino
rates are left unchanged, albeit via novel compensating
mechanisms. The theory of core collapse supernova evolution, and
neutron star cooling, combined with future observations of supernova
neutrinos and multi-wavelength photons from neutron stars has the
potential to probe the inner core of these remarkable compact objects.

\section{Acknowledgments}
I would like to thank George Bertsch, Greg Carter, Mark Alford, David
Kaplan, Jim Lattimer, Jose Pons, Madappa Prakash, Krishna Rajagopal,
Mariusz Sadzikowski, Andrew Steiner, Motoi Tachibana and Frank Wilczek
for enjoyable collaborations.  The materials presented here are the
results of these collaborations over the past four years. Special
thanks are due to Dick Silbar for a careful reading of the
manuscript and several valuable suggestions.  This work is supported
in part by funds provided by the U.S. Department of Energy (D.O.E.)
under cooperative research agreement DF-FC02-94ER40818 and the
D.O.E. contract W-7405-ENG-36.

\end{document}